%% file: main-cr.tex
\documentclass[acmsmall]{acmart}

\usepackage{graphicx}
\usepackage{adjustbox}
\usepackage{microtype}
\usepackage{epstopdf}
\usepackage{flushend}
\usepackage{balance}
\usepackage{colortbl}
\usepackage{xcolor}
\usepackage[dvipsnames]{xcolor}
\usepackage{listings}
\usepackage{booktabs}
\usepackage{nicefrac}
\usepackage{paralist}
\usepackage{caption}
\usepackage{subcaption}
\usepackage{soul}
\usepackage{subfiles}
\usepackage{multirow}
\usepackage{enumitem}
\usepackage{tabularx}
\usepackage{xspace}
\usepackage{verbatim}
\usepackage{graphicx} 
\usepackage{pdfpages}
\usepackage[ruled,linesnumbered,noend]{algorithm2e}
\usepackage{amsmath}
\usepackage[normalem]{ulem}

\usepackage{tikz}

\newcommand*\circled[1]
  {\tikz[baseline=(char.base)]{
    \node[shape=circle, fill=., inner sep=0pt, text=white]
    (char) {\textsf\footnotesize #1};}}

\usepackage{tikz}
\usepackage{booktabs}
\usepackage{xcolor}

\AtBeginDocument{%
  }

\setcopyright{cc}
\setcctype{by}
\acmJournal{PACMMOD}
\acmYear{2026} \acmVolume{4} \acmNumber{3 (SIGMOD)} \acmArticle{134}
\acmMonth{6} \acmDOI{10.1145/3802011}

\newcommand{\siftten}{sift10M\xspace}
\newcommand{\openai}{openai5M\xspace}
\newcommand{\cohere}{cohere10M\xspace}
\newcommand{\texttoimage}{text2image10M\xspace}
\newcommand{\oai}{OpenAI-5M\xspace}

\newcommand{\scann}{ScaNN\xspace}
\newcommand{\acorn}{ACORN\xspace}
\newcommand{\navix}{NaviX\xspace}
\newcommand{\sweeping}{Sweeping\xspace}
\newcommand{\iterativescan}{Iterative Scan\xspace}
\newcommand{\hnswlib}{HNSWLib\xspace}
\newcommand{\pgvector}{PGVector\xspace}
\newcommand{\postgresql}{PostgreSQL\xspace}

\newcommand{\todo}[1]{{\color{red} {\bf #1 }}\normalcolor}

\definecolor{BrickRed}{rgb}{0.8, 0.25, 0.33}

\begin{document}

\title{An In-Depth Study of Filter-Agnostic Vector Search on a PostgreSQL Database System: [Experiments \& Analysis]}

\author{Duo Lu}
\authornote{The work was done while working at Google.}
\email{duo_lu@brown.edu}
\orcid{0009-0002-5901-3079}
\affiliation{%
  \institution{Brown University}
  \city{Providence}
  \state{Rhode Island}
  \country{USA}
}

\author{Helena Caminal}
\email{hcaminal@google.com}
\orcid{0000-0002-2052-8107}
\affiliation{%
  \institution{Google}
  \city{Sunnyvale}
  \state{California}
  \country{USA}
}

\author{Manos Chatzakis}
\authornotemark[1]
\email{manos.chatzaki@gmail.com}
\orcid{0000-0002-9616-6210}
\affiliation{%
  \institution{Université Paris Cité, LIPADE}
  \city{F-75006 Paris}
  \country{France}
}

\author{Yannis Papakonstantinou}
\email{yannispap@google.com}
\orcid{0009-0007-6360-9496}
\affiliation{%
  \institution{Google}
  \city{San Diego}
  \state{California}
  \country{USA}
}

\author{Yannis Chronis}
\email{chronis@ethz.ch}
\orcid{0000-0003-2214-6919}
\affiliation{%
  \institution{ETH Zurich \& Google}
  \city{Zurich}
  \country{Switzerland}
}

\author{Vaibhav Jain}
\email{jainva@google.com}
\orcid{0009-0008-8069-5928}
\affiliation{%
  \institution{Google}
  \city{Bangalore}
  \country{India}
}

\author{Fatma \"{O}zcan}
\email{fozcan@google.com}
\orcid{0000-0002-4418-4724}
\affiliation{%
  \institution{Google}
  \city{Sunnyvale}
  \state{California}
  \country{USA}
}

\renewcommand{\shortauthors}{Duo Lu et al.}



\received{October 2025}
\received[revised]{January 2026}
\received[accepted]{February 2026}

\input{0-abstract}

\maketitle

\input{1-introduction}

\input{2-overview}
\input{4-methods}
\input{5-system-challenges}

\input{6-filters}

\input{7-evaluation-setup}
\input{9-evaluation-analysis}

\input{10-lessons}

\input{11-related-work}
\input{13-conclusion}

\begin{acks}
The authors would like to thank the anonymous reviewers for their constructive feedback.
\end{acks}

\bibliographystyle{ACM-Reference-Format}
\bibliography{citations}

\end{document}

%% file: 0-abstract.tex
\begin{abstract}

Filtered Vector Search (FVS) is critical for supporting semantic search and GenAI applications in modern database systems. However, existing research most often evaluates algorithms in specialized libraries, making optimistic assumptions that do not align with enterprise-grade database systems. Our work challenges this premise by demonstrating that in a production-grade database system, commonly made assumptions do not hold, leading to performance characteristics and algorithmic trade-offs that are fundamentally different from those observed in isolated library settings. This paper presents the first in-depth analysis of filter-agnostic FVS algorithms within a production PostgreSQL-compatible system. We systematically evaluate post-filtering and inline-filtering strategies across a wide range of selectivities and correlations.

Our central finding is that the optimal algorithm is not dictated by the cost of distance computations alone, but that system-level overheads that come from both distance computations and filter operations (like page accesses and data retrieval) play a significant role. We demonstrate that graph-based approaches (such as NaviX/ACORN) can incur prohibitive numbers of filter checks and system-level overheads, compared with clustering-based indexes such as ScaNN, often canceling out their theoretical benefits in real-world database environments.

Ultimately, our findings provide the database community with crucial insights and practical guidelines, demonstrating that the optimal choice for a filter-agnostic FVS algorithm is not absolute, but rather a system-aware decision contingent on the interplay between workload characteristics and the underlying costs of data access in a real-world database architecture. 

\end{abstract}

%% file: 1-introduction.tex
\section{Introduction}
\label{sec:introduction} 

\begin{figure}[t]
\centering
\includegraphics[width=0.55\columnwidth]{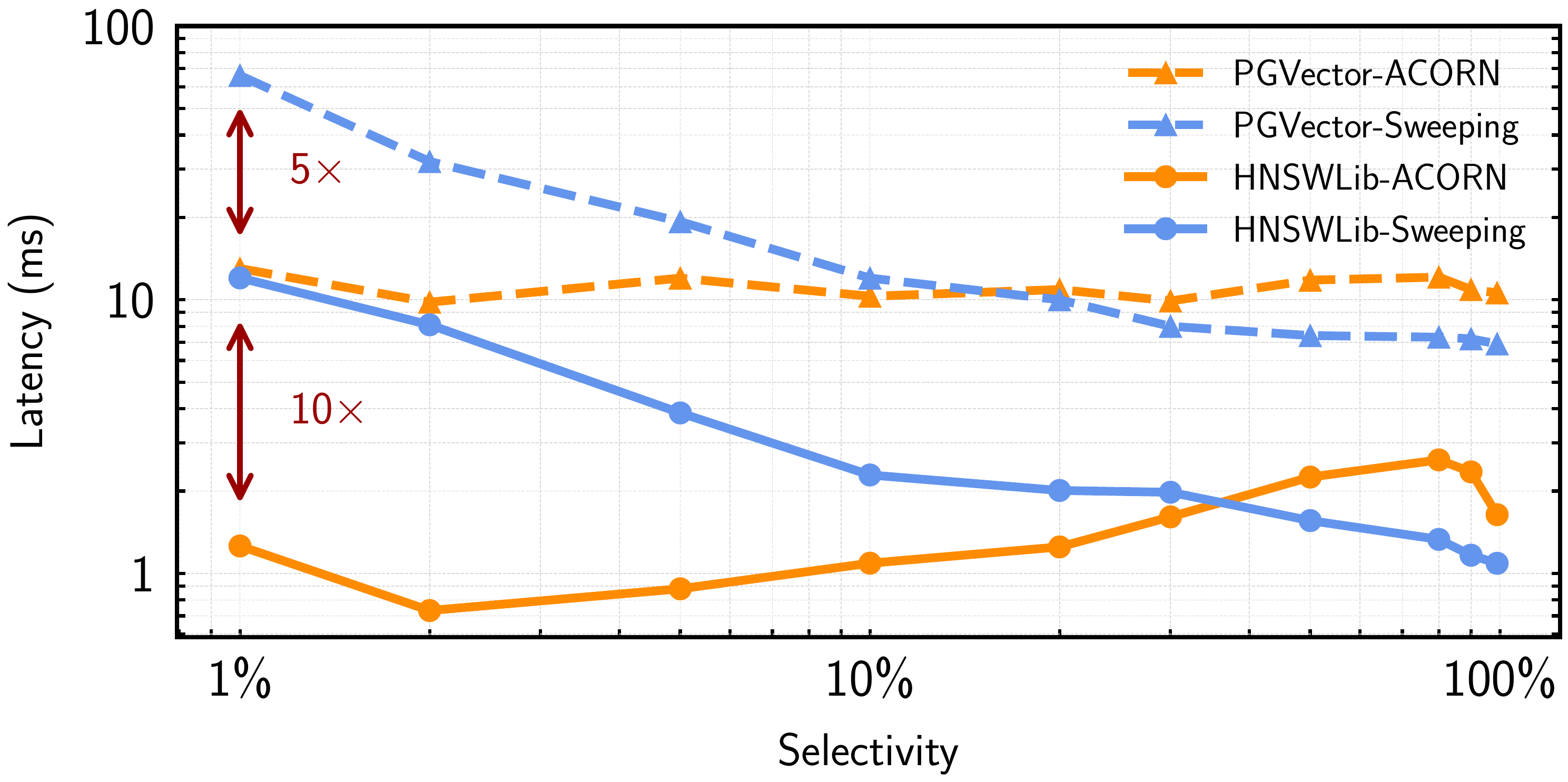}
\caption{
Average latency for 100 filtered vector search queries on OpenAI-1M: \hnswlib (solid) vs. \pgvector (dashed). System latency can be up to 10× higher, and the gap varies with selectivity, motivating an in-depth comparison of FVS algorithms in a real DBMS.
}

\label{fig:motivation-plot}
\end{figure}

Vector embeddings enable semantic search over unstructured data. They underpin modern Generative AI and advanced information retrieval systems~\cite{lewis2020rag, huang2020embeddingretrieval}, and their applications are rapidly proliferating across numerous domains~\cite{gemini2024, chatgpt2024, touvron2023llama}.

Combining semantic search using vector embeddings with structured data filters enhances the quality of semantic query results~\cite{scann4alloydb}, motivating the integration of \textit{Approximate Nearest Neighbor vector Search (ANNS)} in conventional database systems. For example, consider an e-commerce query: a user wants a t-shirt "similar to this image" (vector search), but only those "available in size large" and "in stock locally" (filters on structured data).  Enabling efficient \textit{Filtered Vector Search (FVS)}, in other words, the efficient combination of relational filters, joins, and vector search, is a critical next step in blending traditional database systems with modern AI capabilities.


The baseline execution strategies of FVS are \textit{pre-filtering}, which executes the filtering first, and \textit{post-filtering}, which executes the vector search first. Pre-filtering and post-filtering are well understood and optimal only in some restricted scenarios. Going beyond them, blending filters and vector search has been shown to lead to efficient FVS \cite{patel2024acorn}.
There are multiple ways to order relational operators and vector search, and optimize the vector search itself. 

Most existing research evaluates new FVS execution methods in a vector search library and not in a database system. This setup ignores system overheads and the cost of filter evaluation during the vector index traversal. These costs can dominate over the cost of distance computations, which is believed to be the most costly part of vector search. In the majority of existing studies, performance is often benchmarked solely by minimizing the number of distance computations, even if it requires more filter checks ~\cite{patel2024acorn, fvs-survey}. 


Figure~\ref{fig:filtering-methods} shows significant differences in the performance characteristics of the same vector index traversal algorithms when implemented in a production database (PGVector~\cite{pgvector}) compared to a standalone library (HNSWLib~\cite{hnswlib}). It is expected that the performance of an algorithm will differ when run in a standalone library compared to a fully fledged system. In our example, the total latency on the database system can be up to 10x higher due to system overheads (Figure ~\ref{fig:motivation-plot}), which is expected. However, prior to this work, it was not evident that the performance gap between the two algorithms is not constant across different filter selectivities between the two setups (database and library). Consequently, the selectivity cross-over point—where one search algorithm becomes superior to another can change completely when filtered vector search is integrated into the database system. 

Our observations show that minimizing distance computations alone is an unreliable proxy to improve the end-to-end FVS performance. As detailed in Table \ref{tbl:indexes}, this discrepancy stems from fundamental architectural differences between libraries and fully-fledged database systems. While libraries optimize for direct memory access, packing nodes contiguously and traversing graphs via raw pointers, database systems like \postgresql, introduce necessary layers of abstraction for durability and concurrency. 
In particular, neighbor filtering in a DBMS is rarely a single, lock-free memory dereference: it typically involves multiple engine layers (buffer manager and tuple access) before a candidate can be filtered and scored. As a result, algorithmic choices that look beneficial in libraries (e.g., more aggressive filtering to reduce distance computations) can become regressions once these system costs are accounted for.

It is well-established that the optimal approach for FVS depends on both filter selectivity and vector-predicate correlation~\cite{patel2024acorn}. However, a comprehensive comparison of methods across the full spectrum of selectivity-correlation combinations in a real system is absent. This gap prevents a complete understanding of when and why FVS strategies succeed or fail in a real database system. 

Several existing approaches optimize vector indices for FVS by assuming specific filter patterns ~\cite{gollapudi2023filteredvamana, zuo2024serf, wang2023nhq, ait2025rwalks, Xu2024iRangeGraph, Jiang2025DIGRA, Liang2025UNIFY, peng2025dynamicrfanns, li2025sieve, Cai2024Navigating, Zhang2025Efficient}. These approaches often limit the supported types and number of filter predicates, and thus, are not widely adopted for database workloads with arbitrary filter predicates on many attributes. Therefore, \textit{filter agnosticism}—the ability to efficiently handle arbitrary, unknown filters at query time—is not just a desirable feature but a fundamental requirement for integrating vector search into a versatile database environment. To the best of our knowledge, ANN indices in commercial database systems are filter-agnostic ~\cite{oracle-vector-search-manual, mongodb_atlas_vector_search_filter, elastic_filtered_knn, upreti2025cosmosdb}.


In this work, we evaluate filter-agnostic search strategies across two dominant indexing paradigms: graph-based and clustering-based, within a real database system. We use HNSW~\cite{malkov2020hnsw} to represent graph-based approaches for its widespread adoption ~\cite{wang2023bulletin, aumuller2021roleoflid, li2019approximatebenchmark, echihabi2020hydra2} and ScaNN~\cite{scann}, a performance-optimized clustering-based index. This selection allows us to contrast fundamental system-level trade-offs across index types.
Our experiments use a commercially available \postgresql-compatible system, utilizing various filter selectivities and vector-predicate correlations on four different known vector datasets, ranging from 5 to 10 million vectors. Our goals are to inform a) the development of cost models for FVS queries and b) the design and optimization of vector search methods.

\noindent
We summarize our contributions:
\begin{itemize} [wide=0pt, nosep]
    \item We present filter vector search (with ANNS) evaluations conducted in a real DB system, focusing on filter-agnostic methods.
    \item We design a novel, dataset-agnostic workload generator that simulates workloads of different filter correlations and selectivities.
    \item We present guidelines and insights of filter-agnostic vector search algorithms, based on our observations.
\end{itemize}

%% file: 2-overview.tex
\section{Background on Filtered Vector Search}






\begin{table}[t]
\caption{
System breakdown for HNSW: architectural differences between standalone libraries and database systems.
}
\label{tbl:indexes}
\centering
\small

\begin{tabular}
{@{}m{1.45cm}m{3.8cm}m{3.8cm}m{3.8cm}@{}}
\toprule
\textbf{Arch. Aspect} & \textbf{Library} & \textbf{System (PGVector)} & \textbf{System overheads} \\ \midrule
\textbf{Index Storage} 
& \textbf{Compact representation:} Neighbors stored as raw pointers. Nodes are packed contiguously with no alignment constraints.
& \textbf{Tuple-based storage:} Each neighbor reference is a 6-byte TID. Nodes are stored as tuples, with header overhead, in 8KB pages.
& \textbf{Space amplification:} Page-level overhead (headers, alignment) reduces effective storage density, impacting cache efficiency. \\

\rowcolor[HTML]{EFEFEF} 
\textbf{Node Access}  
& \textbf{Direct deference:} The neighbor access is a CPU-level memory dereference.
& \textbf{Scattered page access:} The neighbors are stored as TIDs pointing to arbitrary pages.
& \textbf{Random I/O amplification:} Each hop multiplies page overhead by neighbor fanout. \\

\textbf{Filter Evaluation} 
& \textbf{Unified identifier:} A single ID/offset indexes both vector data and attribute data in memory.
& \textbf{Indirection:} Index page $\rightarrow$ HeapTID $\rightarrow$ Heap page. Filter columns reside in the heap. Evaluating predicates needs HeapTID.
& \textbf{Double lookup:} Every filtered candidate requires: (1) index access for vector+HeapTID, (2) heap access for filter columns. \\

\rowcolor[HTML]{EFEFEF} 
\textbf{Parallelism}  
& \textbf{Flexible:} Intra-query parallelism via OpenMP, \texttt{std::thread}, or SIMD.
& \textbf{Constrained:} One query per connection. No batch query primitive.
& \textbf{Arch. limit:} Executor/storage hinders fine-grained parallelism and batching. \\
\bottomrule
\end{tabular}
\end{table}

Databases blend relational operators with vector search. While database systems have long studied query planning and optimization across multiple operators, vector search poses a challenge. Most vector search use cases tolerate approximation in solution retrieval, enabling Approximate Nearest Neighbor Search (ANNS) via vector search indexes (e.g., HNSW~\cite{malkov2020hnsw}, ScaNN~\cite{scann}). The goal is to significantly reduce the data processed while maximizing recall. 


\subsection{Execution Strategies for FVS Queries}

\begin{figure}[h!]
\centering
\includegraphics[width=0.52\columnwidth]{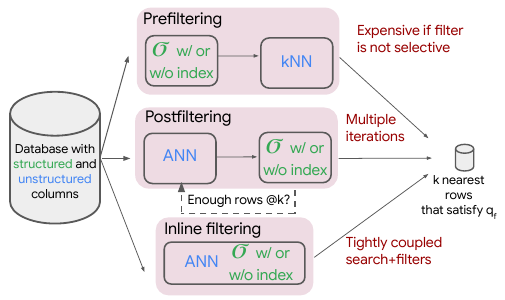}
\caption{Execution strategies for an FVS query.}
\label{fig:filtering-methods}
\end{figure}

\noindent
Based on the execution order of filter evaluation and vector search operators, we classify the strategies as follows (Figure~\ref{fig:filtering-methods}):
 \begin{itemize} [wide=0pt, nosep]
     \item Pre-filtering: queries with extremely selective filters should filter first, then apply exact nearest neighbor (KNN) to surviving tuples.
     \item Post-filtering: if the filters are not selective, we should probe the ANN index, and then filter the result. 
     \item Inline-filtering: tightly-coupled searching and filtering.
\end{itemize}

\noindent
Note that to maintain an acceptable recall in post-filtering cases, we ought to retrieve $k+\delta$ tuples from the vector index, to guarantee that after the filter is applied, there are at least $k$ tuples left. Predicting the optimal $\delta$ is non-trivial; however, both over- and under-provisioning have performance and quality implications. Alternatively, one may go back to probing the index, retrieving another round of NN points. We consider this last option (also known as \emph{iterative post-filtering} or \emph{iterative scan}) to belong to the inline-filtering family of execution strategies, since it merges filter evaluation and vector search.

In this paper, we focus on filter-agnostic inline filtering vector search methods. Filter-agnostic methods are better suited to database workloads that are dynamic and complex, and for which filters are not known a priori. Pre-filtering and inline filtering are complementary. After evaluating the filter predicates, query execution can choose the pre- or inline-filtering depending on the selectivity. Pre-filtering is faster when filters are very selective~\footnote{In some experiments, we have observed selectivity ranges below 0.01\% benefiting from pre-filtering.}. Pre-filtering, as an execution method, is straightforward for an optimizer to estimate its cost.
In contrast, the performance of inline filtering is not well understood. We do not explicitly evaluate post-filtering because the iterative scan method subsumes it: once sufficient solutions are found in the first round of iterative scan, iterative scan and post-filtering are equivalent. The adaptability of iterative scan to the selectivity has made it a de-facto choice for FVS in \pgvector systems. In conclusion, this study focuses on filter-agnostic inline filtering algorithms and their implementation in a real-world system.


\subsection{Inline Filter-Agnostic Approaches}

Filter-agnostic FVS decouples index construction from structured filters~\cite{fvs-survey}, offering three advantages: \circled{1} \textbf{Arbitrary Filter Support:} Unlike specialized indexes limited to low-cardinality predicates, these handle any SQL filter—from equality to complex regex—without index modifications. \circled{2} \textbf{Attribute-Blind Construction:} By ignoring attribute data during indexing, the structure remains valid and performant despite schema changes. \circled{3} \textbf{Update Resilience:} Modifications to attribute values never degrade the index or necessitate costly rebuilding, making it ideal for dynamic environments.

\label{sec:overview}

%% file: 4-methods.tex
\subsection{Filter-Agnostic FVS Algorithms}
\label{sec:methods}

In our study, we focus on the two dominant paradigms for ANNS indexing: graph-based and clustering-based indexes. Specifically, we utilize Hierarchical Navigable Small World (HNSW)~\cite{malkov2020hnsw} and \scann~\cite{scann}, for their wide adoption.

\subsubsection{Unfiltered HNSW Search} 
The HNSW search algorithm consists of two phases:
\noindent \textbf{(i) Zoom-in Phase:} Starting at a top-layer entry point, the algorithm greedily traverses each layer to find the node closest to the query, using it as the entry point for the subsequent layer until reaching the base layer (layer 0).
\noindent \textbf{(ii) Base Layer Search:} To find the $k$ nearest neighbors, the search explores layer 0 using a min-priority queue of unvisited candidates ($C$), a max-priority queue for the top-$ef$ results ($W$), and a set of visited nodes ($V$). At each step, the algorithm:
\begin{itemize}[wide=0pt, nosep]
\item Pops the top candidate from $C$ and evaluates unvisited neighbors.
\item Adds neighbors closer than the farthest node in $W$ to both $C$ and $W$, maintaining $|W| \leq ef\_search$.
\item \textit{Stops when the top candidate in $C$ is farther than $W$'s farthest}.
\end{itemize}
\noindent The final result set contains the $k$ closest vectors from $W$.

\begin{figure}[h!]
\centering
\includegraphics[width=0.675\columnwidth]{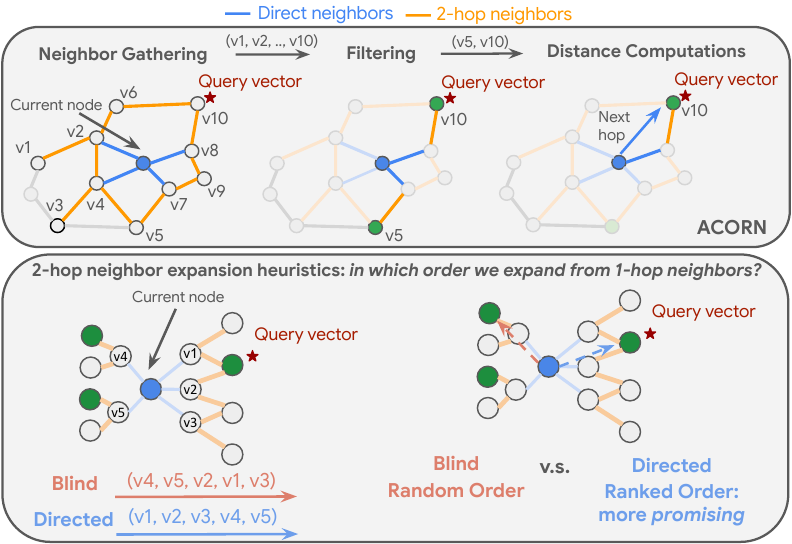}
\caption{Filter-first execution methods for an FVS query.}
\label{fig:filter-first}
\end{figure}

\begin{figure}[h!]
\centering
\includegraphics[width=0.675\columnwidth]{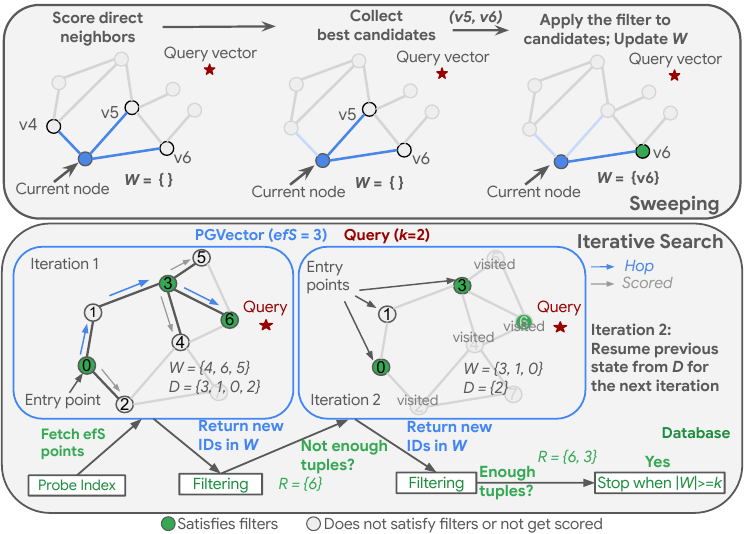}
\caption{Traversal-first execution methods for an FVS query.}
\label{fig:traversal-first}
\end{figure}

\begin{figure}[h!]
\centering
\includegraphics[width=0.675\columnwidth]{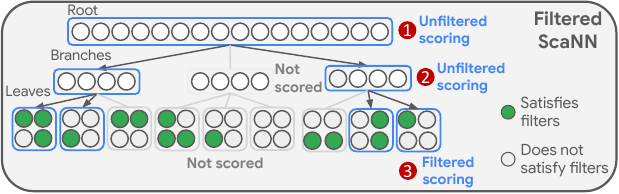}
\caption{
Execution methods for an FVS query on \scann.
}
\label{fig:scann-filtering-methods}
\end{figure}

\subsubsection{Graph-based Filtered Search Algorithms}
Integrating filter predicates into the graph search algorithms introduces a fundamental design choice: when should we check if a node satisfies the filter predicates along the graph traversal? We categorize the evaluated three algorithms as:

\noindent
\textbf{Filter-first (\navix/\acorn)} ~\cite{sehgal2025navix, patel2024acorn}: 
During traversal, it first checks if a node satisfies filters and, if it does not, it skips performing the distance computation.


\noindent
\textbf{Traversal-first (\textit{\sweeping}) ~\cite{sweeping-wea}:} performs graph navigation using the original unfiltered structure, checking if a vector satisfies the filter only before moving candidates to the result priority queue. \noindent
\textbf{\iterativescan:} ~\cite{pgvector} uses the index to produce candidate batches that are filtered after the graph traversal is complete.

\noindent
Next, we deep dive into each of these methods.




\subsubsection{ACORN}
\label{sec:acorn-imple}
Building on the HNSW index, \acorn was designed to address the inflexibility of filter-aware indexes.

\noindent
\textbf{Predicate Subgraph Traversal.}
At query time, the algorithm traverses the subgraph formed only by the nodes that pass the query's predicate (Figure~\ref{fig:filter-first}-top). By ignoring neighbors that do not satisfy the filter, the search effectively steps across the graph to find paths within the predicate-specific subgraph. 




\noindent
Arbitrary predicate subgraphs can be too sparse or disconnected for efficient navigation. To maintain connectivity, ACORN employs \textit{run-time 2-hop neighbor expansion} (Figure~\ref{fig:filter-first}-top): at each traversal step, it gathers 1-hop and 2-hop neighbors, filters them, and computes distances only for qualifying nodes to determine the next hop. ACORN-1 relies solely on this search-time expansion, while ACORN-$\gamma$ also densifies the graph during construction, using compression to manage size.

\subsubsection{\navix}
\label{sec:navix-algo}

The \navix algorithm design builds upon the \acorn-1 baseline but introduces a dynamic execution layer to optimize traversal under varying filter selectivities. The method defines three distinct heuristics:
\textbf{Blind (Figure~\ref{fig:filter-first}-bottom-left):} This strategy is generally equivalent to \acorn's 2-hop neighbor expansion. But it explores 1-hop neighbors first, followed by an expansion into 2-hop connections.
\textbf{Directed (Figure~\ref{fig:filter-first}-bottom-right):} A guided traversal that, when considering filtering 2-hop neighbors, prioritizes 2-hop neighbors from the closest (highest-ranked) 1-hop nodes, rather than exploring them indiscriminately.
\textbf{Onehop-s:} A standard greedy search performed exclusively on filtered 1-hop neighbors.
\navix employs an \textit{adaptive-local} mechanism to select the most appropriate heuristic at every traversal step based on the observed filter selectivity.

\subsubsection{\sweeping} 
\label{sec:sweeping-imple}
\sweeping \cite{sweeping-wea} populates the result set ($W$) based on proximity, as the non-filtered traversal, but conditionally to the points satisfying the filter (Figure ~\ref{fig:traversal-first}). 

\subsubsection{\iterativescan}
\label{subsec:iterativescan}

It implements a "resumable" post-filtering strategy by decoupling 
the HNSW index traversal from the database (filter) executor. 
At each iteration, the algorithm maintains a \textit{discarded candidate queue} ($D$) containing visited nodes that were conceptually "paused" --- either because (i) they were popped from the result set ($W$) or (ii) were seen but not yet explored (Figure~\ref{fig:traversal-first}-bottom).
If the initial batch yields fewer than $k$ results after applying the filters, 
the search resumes execution by filling up its entry points and queues from the state preserved in $D$. 

\subsubsection{Filtered \scann}
\label{sec:scann-algo}
In a three-level tree, all root centroids are scored and used to select the closest branch candidates to the query vector (Fig.~\ref{fig:scann-filtering-methods}-\circled{1}). Next, all these branch centroids are scored, similarly to the root centroids, to decide the best leaf candidates according to distance only (Fig.~\ref{fig:scann-filtering-methods}-\circled{2}). Finally, in a filtered search, only points in leaves satisfying filters are fetched and scored (Fig.~\ref{fig:scann-filtering-methods}-\circled{3}).




%% file: 5-system-challenges.tex
\section{FVS Algorithms in \postgresql Database}

\begin{figure}[h!]
\centering
\includegraphics[width=\columnwidth]{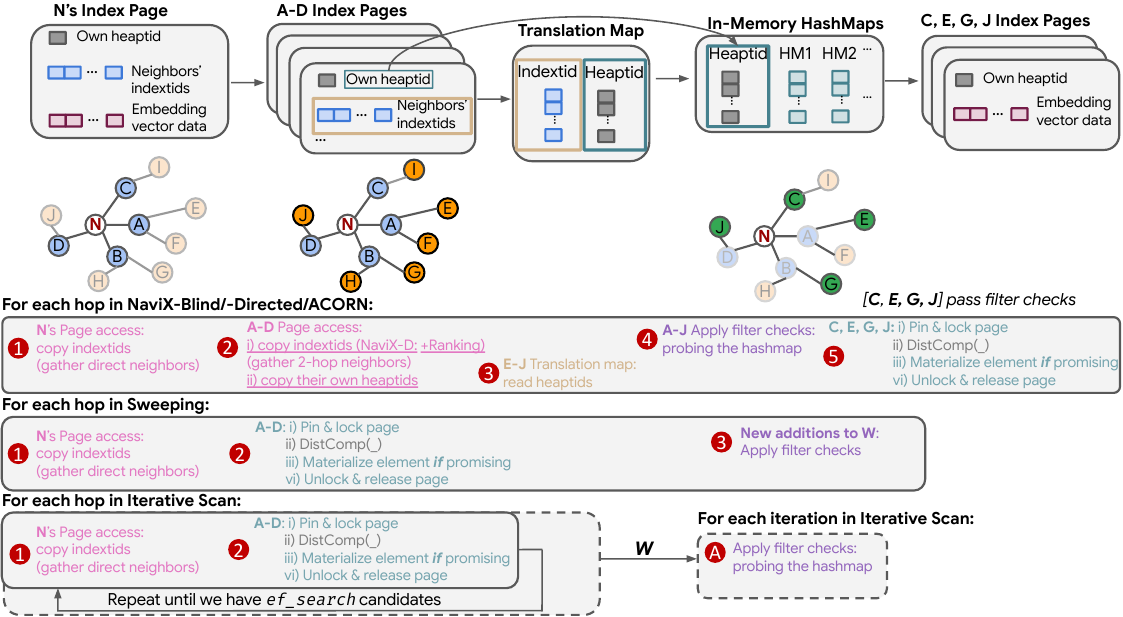}
\caption{
\navix, \acorn, \sweeping, and \iterativescan implementations in \pgvector.
}
\label{fig:acorn-pgvector}
\end{figure}

\begin{figure}[h!]
\centering
\includegraphics[width=0.65\columnwidth]{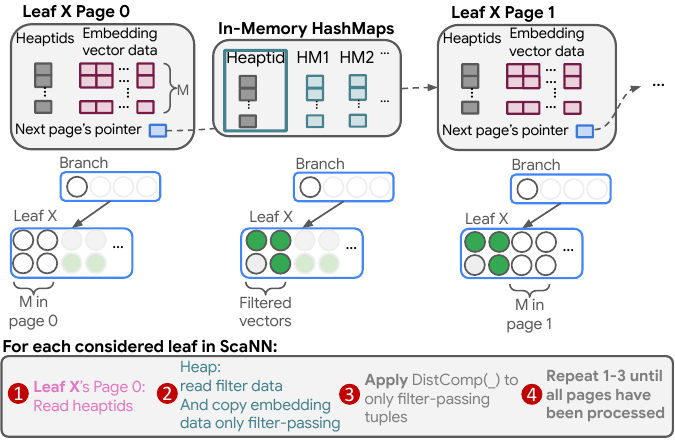}
\caption{
Filtered \scann implementation in the system.
}
\label{fig:scann-pgvector}
\end{figure}

In this section, we describe the system details that implement the previously described inline filtering methods on a \postgresql (PG) commercial-grade system.

\subsection{Graph-based Filter-First Methods}
\label{subsec:acorn-system}

\noindent \textbf{Physical Layout.}
In PG, data is persisted in fixed-size (of, by default, 8KB) pages. 
Index Pages in \texttt{pgvector} contain information about: 
(i) the node element: high-dimensional vector data and a pointer to the heap tuple,
and (ii) information about its neighbors: pointers to neighbor pages. 


\noindent \textbf{The implications of the Page Limit.}
\noindent
The current \texttt {pgvector} enforces a constraint for the neighbor information of a node to fit within a single page. This creates a hard structural limit on the graph topology:
\begin{equation}(L_{max} + 2) \cdot M \cdot S_{ptr} \le S_{page}\end{equation}
where $M$ is the maximum number of connections per layer, $L_{max}$ is the maximum layer height, $S_{ptr}$ is the size of a neighbor pointer, 
and the constant $2$ accounts for the base layer allowing $2M$ connections (standard HNSW specification).
This storage constraint creates a zero-sum trade-off between the graph's connectivity ($M$) and its navigability ($L_{max}$).
For example, with a standard configuration where $M=40$,
$L_{max} \approx 31$
If we attempt to implement ACORN-$\gamma$ with $\gamma=2$ by simply doubling $M$ to 80, the maximum hierarchy depth $L_{max}$ drastically drops to $\approx 14$. For larger $M$, the graph rapidly collapses to merely 2 or 3 layers.

\noindent \textbf{TOASTing (\postgresql's \textit{The Oversized-Attribute Storage Technique}) the neighbor lists is not an option}
as visiting any node would trigger multiple random I/O requests just to materialize the node's neighbors and degrade the query latency significantly.

\noindent
\textbf{System Level Costs.}
A crucial and often overlooked factor in database system performance is the page retrieval and locking overhead associated with page access. In a row-based system like \postgresql, a node's index page stores two types of identifiers: the $indextids$ of its neighbors (pointing to other pages within the index structure) and the $heaptid$ of the node itself (pointing to the row in the database table).
Consequently, complex traversal algorithms that perform aggressive neighbor expansion, such as ACORN and the \textit{Blind} or \textit{Directed} heuristics of \navix, incur a significant number of page accesses.
To evaluate a candidate, the system cannot simply jump to the data; it must first fetch the candidate's index page to retrieve its $heaptid$.
For strategies relying on two-hop expansion (e.g., \acorn and \navix-Blind/Directed), this cost compounds geometrically.
For a single node, the system must first fetch its index page to retrieve the $indextids$ of its $M$ neighbors (Figure~\ref{fig:acorn-pgvector}-\circled{1}). Then, it must fetch each of those $M$ neighbor pages to retrieve their $heaptids$ for filtering and to discover the $indextids$ of the two-hop neighbors (Figure~\ref{fig:acorn-pgvector}-\circled{2}). Finally, to filter the (up to) $M \times M$ two-hop neighbors, it must access their respective index pages to retrieve their $heaptids$ (Figure~\ref{fig:acorn-pgvector}-\circled{4}).
In the worst case, this results in $1 + M + (M \times M)$ page accesses, each requiring a buffer pool lookup and a page lock.

\noindent
\textbf{Our Optimizations.}
Porting \navix/\acorn into a production row-oriented relational database necessitates adapting its design to the system's architectural constraints. A direct, naive implementation would be unacceptably slow due to the costs described previously.
We introduced two critical optimizations, 
\textit{(i) A pre-computed indextid-to-heaptid Translation Map}, implemented as an in-memory hash map (Figure~\ref{fig:acorn-pgvector}-\navix/\acorn's \circled{3}. This map is generated during index build time. At query time, this map is loaded and used to perform the indextid to heaptid translation in-memory (Figure~\ref{fig:acorn-pgvector}, Step 3). This optimization effectively eliminates an entire class of 2-hop neighbors' page accesses for a \navix-Blind/\navix-Directed/\acorn search step. \textit{(ii) An adaptive 2-hop neighbor expansion (Hardening \acorn).}
We implemented a conditional strategy: if a 1-hop neighbor passes the filter, we skip the expensive 2-hop expansion for that branch. While this simple adaptive heuristic differs from \navix's complex navigation logic, it is highly effective in the database system. It drastically reduces page access overheads in high-selectivity regions, ensuring \acorn serves as a competitive, optimized baseline.

\subsection{Graph-based Traversal-First Methods}

\noindent
\textbf{\sweeping (Inline Filtering).} We implemented \sweeping ~\cite{weviate} by modifying the internal HNSW loop to perform checks before adding candidates to the result set. 

\noindent
\textbf{\iterativescan.} We utilize the implementation in \pgvector 0.8.0. From a DBMS integration standpoint, \iterativescan behaves fundamentally as a post-filtering method, following the logic described in Section~\ref{sec:methods}.

\noindent


\subsection{Filtered \scann}
As described in Section~\ref{sec:scann-algo}, the ScaNN-based filtered vector search performs the filtering at the leaf level only. The nature of cluster-based indexes allows for the implementation to be much more compact in size and simpler in number of steps: each leaf packs as many vectors as they fit in a single page (8KB) and maintains a linked list of pages of the same leaf (Figure~\ref{fig:scann-pgvector}, top). This allows for: a) simpler access to neighbors, and b) easier SIMD scoring. At the system level, once a leaf has been selected for scoring, the first page is accessed and the heaptids of the vectors in the leaf are used to read the in-memory hashmaps used for filtering (Fig.~\ref{fig:scann-pgvector}-\circled{1} and \circled{2}). At last, only tuples passing filters are scored using SIMD (Fig.~\ref{fig:scann-pgvector}-\circled{3}). 

\subsection{Steps in a Search Algorithm in a System}
\label{subsec:search-steps}
\subsubsection{Graph-based Indexes}
\label{sec:graph-steps}
\noindent
In summary, the types of actions a system may execute across search algorithms are as follows (Figure~\ref{fig:acorn-pgvector}):

\noindent
\circled{1} \textbf{\uline{Accessing one-Hop Neighbors:}} Access the current node's index page to read its direct neighbors' indextids (BlockNumber and OffsetNumber pair within the index file).

\noindent
\circled{2} \textbf{\uline{Gathering two-Hop Neighbors:}} \navix-Blind/\acorn: access the $M$ direct neighbors index pages to gather their respective $M$ neighbors' (two-hop neighbors) indextids. \navix-Directed: access $M$ direct neighbors index pages to score \& rank all direct neighbors.

\noindent
\circled{3} \textbf{\uline{Translation for two-hop Neighbors:}}  \navix-Blind / \navix-Directed / \acorn: access the Translation Map (our optimization) to fetch the two-hop neighbors' heaptids.

\noindent
\textbf{\uline{Apply Filter Checks}} (\circled{4} in \navix/\acorn or \circled{3} in \sweeping or \circled{A} in \iterativescan): Access the in-memory hashmaps (pre-generated, as we describe in more detail in Section~\ref{sec:exp_setup}).

\noindent
\textbf{\uline{Vector Retrieval \& Computing Distance:}} (\circled{5} in \navix/\acorn (part of \circled{2} in the \navix-Directed ranking phrase) or \circled{2} in \sweeping and \iterativescan). This step scores a potential candidate node through a sequence of optimized operations designed to minimize memory overhead and unnecessary work. The process is not a single action but a conditional workflow, including page access, tuple retrieval, distance computation, conditional materialization, and resource release. Further details and profiling results are provided in Section ~\ref{sec:profiling}.

\noindent
Additionally, a Page Access involves multiple steps, including:
    (1) Page pinning in the memory buffer to prevent eviction and acquiring a shared lock to ensure a consistent read.
    (2) Reading the data directly from the shared page buffer. 
    (3) The data for any given tuple, including its vector, resides on a shared buffer page that is only protected by a short-term lock. However, the search algorithm needs to maintain a working set of the best candidates found so far (the set $W$) for the entire duration of the search on a given layer. To safely use a candidate's data across multiple comparisons --- long after its original page lock is released --- it must be copied into a private, query-local memory context. This materialization involves memory allocation ($palloc$) and copying the potentially large vector, making it an expensive operation.
    (4) Finally, the lock on the buffer is released, making the page available for other transactions.


\subsubsection{\scann Indexes}
\scann's index traversal is simpler to support in a PG system, compared to graphs. The root and branches traversal relies on walking the linked lists by accessing pages of the same cluster sequentially. Since there is no need to keep explicit neighbor information, the page can be filled with just multiple embedding vectors' data and the pointer to the next page.
The leaves follow a similar flow to the one described in Section~\ref{sec:graph-steps}'s \textbf{\uline{Vector Retrieval \& Computing Distance}} but in a batched way since a page references multiple tids for the filters. Only the embedding vectors of nodes passing filters are scored against the query vector.

\label{sec:system challenges}

%% file: 6-filters.tex
\section{Filtered Vector Workload Generator}

\subsection{The need for exhaustive correlation studies}
\label{sec:filters}
Filters eliminate rows and thus the vectors that should be considered by the vector search of an FVS query. Filter agnostic vector indices are built on the unfiltered data, filters can make the index traversal significantly more difficult compared to the same search without filters. Concretely, filter selectivity and vector-predicate correlation change the distribution of the vectors satisfying the filters in the vector space, and change the vector search hardness (the effort to achieve the same recall).

Existing benchmarks based on real queries and datasets do not control selectivity and correlation simultaneously~\cite{thomee2016yfcc100m, wikiann}, whereas synthetic benchmarks do not exhaustively evaluate queries with varying correlations ~\cite{wang2023nhq, sehgal2025navix, Cai2024Navigating, ait2025rwalks, gollapudi2023filteredvamana, zuo2024serf, pennington2014glove, jegou2011Sift, simhadri2022yantti, openai5M, Cohere10M}. 

To address this limitation, we developed a sophisticated filtered vector search workload generator that can be used to evaluate the behavior and performance of filter-agnostic FVS algorithms. Our goal is to evaluate the vector index search component of filter-agnostic inline-filtering vector search methods using an execution strategy that first evaluates the filters and then passes the filter results as a bitmap to the vector index search.

Given a vector dataset, a vector query, a selectivity, and a correlation type, our generator produces a set of row IDs that satisfy the filters and meet the selectivity and correlation type specification. The list of rowIDs produced for each query and query specification simulates the result of evaluating the filter predicates, without the need to artificially generate structured data and filter predicates to achieve specific selectivity and correlation specifications. From the perspective of vector index search, this setup can simulate queries with a single complex predicate or multiple complex predicates.


\subsection{Implementation}
\label{sec:filter-workload-generation}
\begin{figure}[]
\centering
\includegraphics[width=0.65\columnwidth,trim={.8cm 10.2cm 16.25cm 0.4cm},clip]{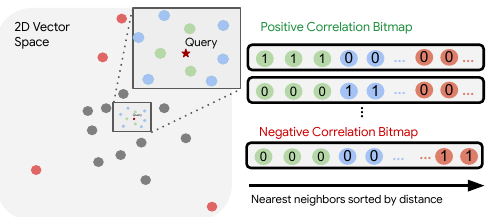}
\caption{Illustration of the filter generator methodology. A Positive Correlation workload is then generated by preferentially sampling vectors from the closest neighbors, conversely, for the Negative Correlation.}
\label{fig:filter-generation}
\end{figure}

First, for every query, the generator calculates the distances of all the database vectors from the query vector. For each query, the database vectors are sorted by increasing distance and stored in an array.  For each query selectivity and correlation combination, we generate the query result by sampling row IDs from the sorted array. The selectivity parameter controls the number of rowIDs sampled, and the correlation type determines the sampling method.
\noindent
Figure~\ref{fig:filter-generation} illustrates how our generator takes into account the correlation type during sampling. For example, a positive vector-predicate correlation ~\cite{patel2024acorn} means that the vectors closest to the query vectors have a higher probability of satisfying the filters. We consider four correlation types. In practice, we observe and expect most queries to have a positive correlation; therefore, we created three positive correlation types and one negative correlation type.

\noindent\textbf{High Positive Correlation:}
For a high positive correlation, our generator samples from only the first third of the sorted array, in other words, the closest vectors to the query. In Figure~\ref{fig:filter-generation}, that would be the green vectors. We perform biased sampling (by applying softmax) based on the distance of each vector to the query vector, to ensure that the vectors closer to the query have a higher probability of being selected.

\noindent\textbf{Medium Positive Correlation:}
Generating workloads with medium positive correlation follows the same process as the high positive correlation.
The key difference is that for medium positive correlation, the generator considers the first half of the array for every query, rather than only the first third.

\noindent\textbf{Low Positive Correlation:}
Generating workloads with low positive correlation follows the other positive correlation cases. However, for low positive correlation, the workload generator samples from the entire sorted array using the same biased sampling method.

\noindent\textbf{Negative Correlation:}
The negative correlation first multiplies all distances of the sorted array by $-1$ and then proceeds identically to the low positive correlation case. 

\noindent\textbf{No Correlation:}
Non-correlated query workloads can be created by simply performing a random sampling over the dataset vectors.

%% file: 7-evaluation-setup.tex
\section{Experimental Setup}
\label{sec:exp_setup}

\begin{table}[t]
\centering
\caption{Comparison and characteristics of benchmark datasets.}
\label{tab:dataset_char}
\footnotesize 
\setlength{\tabcolsep}{7.5pt}
\begin{tabular}{lrrclll}
\toprule
\textbf{Dataset} & \textbf{\# Vec} & \textbf{Dims} & \textbf{Metric} & \textbf{Source} & \textbf{Dist.-Filt. Rel. Cost} & \textbf{LID/LRC} \\
\midrule
\siftten~\cite{jegou2011Sift}    & 10M & 128  & L2 & Image Descriptors & 0.8x & 19.11/0.82 \\
\openai~\cite{openai5M}          & 5M  & 1536 & IP & Text Embeddings   & 6x   & 33.34/0.97 \\
\cohere~\cite{Cohere10M}         & 10M & 768  & L2 & Text Embeddings   & 3.3x & 38.63/0.97 \\
\texttoimage~\cite{simhadri2022yantti} & 10M & 200  & L2 & Multimodal Emb.   & 1.3x & 52.47/0.98 \\
\bottomrule
\end{tabular}
\end{table}

\begin{table}[t]
\centering
\caption{Index build time and size comparison.}
\label{tab:index_perf}
\small
\begin{tabular}{lrrrr}
\toprule
& \multicolumn{2}{c}{\textbf{Time (min)}} & \multicolumn{2}{c}{\textbf{Size (GB)}} \\
\cmidrule(lr){2-3} \cmidrule(lr){4-5}
\textbf{Dataset} & \textbf{HNSW} & \textbf{ScaNN} & \textbf{HNSW} & \textbf{ScaNN} \\
\midrule
\siftten    & 24.80 & 1.87  & 9.58  & 1.73 \\
\openai     & 49.10 & 10.36 & 38.15 & 1.89 \\
\cohere     & 72.83 & 18.24 & 38.14 & 1.95 \\
\texttoimage& 20.35 & 2.84  & 12.72 & 2.35 \\
\bottomrule
\end{tabular}
\end{table}

\noindent \textbf{System.} All experiments were conducted on a server equipped with two 32-core AMD EPYC 7B13 CPUs (64 hyperthreads) and 240 GB of RAM, running Debian GNU/Linux (kernel 6.12.32).
Our benchmark is implemented in a commercial-grade PostgreSQL-compatible database system using the \pgvector~\cite{pgvector} extension, and includes the \scann extension evaluated in the same PostgreSQL database environment.
Since \pgvector does not support intra-query parallelism, we evaluate throughput under load using workload-level parallelism: a client-side driver issues queries over multiple concurrent database connections.
We use 16 client processes to execute queries from the workload in parallel.

\noindent \textbf{Index Configurations.}
We next describe the index configurations and tuning methodology used throughout our evaluations.

\begin{table}[h]
\centering
\small
\caption{QPS speedup and reduction in index size and build time of 1,000 unfiltered vector search queries in PGVector-HNSW with 16 threads (Microbenchmark 95\% Recall@10). }
\label{tab:hnsw_quant_speedup}
\begin{tabular}{llccc}
\toprule
\textbf{Dataset} & \textbf{Method} & \textbf{QPS} & \textbf{Index Size} & \textbf{Build Time} \\
\midrule
\multirow{2}{*}{sift10M} & Halfvec & 0.97$\times$ & {1.34$\times$} & 1.01$\times$ \\
 & BQ+rerank & --* & {2.00$\times$} & 1.02$\times$ \\
\midrule
\multirow{2}{*}{openai5M} & Halfvec & 1.03$\times$ & {2.00$\times$} & 1.11$\times$ \\
 & BQ+rerank & 0.75$\times$ & {11.63$\times$} & {2.86$\times$} \\
\midrule
\multirow{2}{*}{cohere10M} & Halfvec & 1.04$\times$ & {2.00$\times$} & 1.24$\times$ \\
 & BQ+rerank & 0.96$\times$ & {11.83$\times$} & {2.33$\times$} \\
\midrule
\multirow{2}{*}{text2image10M} & Halfvec & 0.99$\times$ & {1.53$\times$} & 1.14$\times$ \\
 & BQ+rerank & --* & {2.55$\times$} & 1.17$\times$ \\
\bottomrule
\multicolumn{5}{l}{\footnotesize{*Unable to reach 95\% recall within reasonable search parameters.}}
\end{tabular}
\end{table}

\begin{table}[h]
\centering
\small
\caption{Latency speedup of different settings w.r.t. non-quantized non-PCAed ScaNN on 95\% recall@10 with 16 threads (no correlation). Columns show increasing selectivity. Quantization uses SQ8.}
\label{tab:scann_ablation}
\newcolumntype{C}[1]{>{\centering\arraybackslash}p{#1}}
\begin{tabular}{llC{0.75cm}C{0.75cm}C{0.75cm}C{0.75cm}C{0.75cm}}
\toprule
\textbf{Dataset} & \textbf{Method} & \textbf{1\% sel.} & \textbf{5\% sel.} & \textbf{20\% sel.} & \textbf{50\% sel.} & \textbf{80\% sel.} \\
\midrule
\multirow{1}{*}{sift10M} & Quant. & 1.40$\times$ & 1.42$\times$ & 1.56$\times$ & 1.49$\times$ & 1.58$\times$ \\
\midrule
\multirow{2}{*}{openai5M} & PCA & 29.39$\times$ & 5.79$\times$ & 3.92$\times$ & 3.22$\times$ & 3.62$\times$ \\
 & PCA+Quant. & 50.21$\times$ & 8.61$\times$ & 6.06$\times$ & 5.02$\times$ & 5.17$\times$ \\
\midrule
\multirow{2}{*}{cohere10M} & PCA & 5.24$\times$ & 4.47$\times$ & 3.48$\times$ & 3.76$\times$ & 3.83$\times$ \\
 & PCA+Quant. & 9.60$\times$ & 7.27$\times$ & 6.55$\times$ & 6.48$\times$ & 5.52$\times$ \\
\midrule
\multirow{1}{*}{text2image10M} & Quant. & 2.17$\times$ & 1.51$\times$ & 1.45$\times$ & 1.87$\times$ & 1.79$\times$ \\
\bottomrule
\multicolumn{7}{l}{\footnotesize{*sift10M and text2image10M are low-dimensional datasets. PCA is not applicable.}} \\
\multicolumn{7}{l}{\footnotesize{*PCA vector dimension reduction (openai5M: 1536$\to$193, cohere10M: 768$\to$157).}}
\end{tabular}
\end{table}

\noindent \uline{Quantization.}
A key difference among the evaluated index families lies in their reliance on quantization.
\textbf{(i) HNSW in \pgvector:} As shown in Table~\ref{tab:hnsw_quant_speedup}, we evaluated \pgvector's available, off-the-shelf quantization options for HNSW---scalar quantization (\texttt{halfvec}/SQ16) and binary quantization with re-ranking---using a microbenchmark on \emph{unfiltered} search at 95\% recall@10.
Across datasets, quantization substantially reduces index footprint and build time (up to $\sim$12$\times$ and $\sim$3$\times$, respectively), but yields \emph{no consistent QPS improvement} for recall (ranging from 0.75$\times$ to 1.04$\times$; in some cases failing to reach 95\% recall within reasonable search parameters). This observation is consistent with prior reports from \pgvector community ~\cite{katz2024pgvectorquant}. 
This behavior is expected because \pgvector-HNSW remains dominated by \emph{random page accesses}: higher page density reduces bytes moved related to embedding data, but does not reduce the number of neighbor-related page fetches and buffer-manager interactions on the critical path.
Therefore, since quantization does not improve latency/QPS at our target recall, we report HNSW results using the unquantized index and focus our tuning on run-time traversal parameters.
\textbf{(ii) \scann:}
In contrast, the clustering-based \scann design is inherently compatible with quantization: it scans leaves/clusters sequentially and can leverage dense, quantized representations to achieve higher throughput (e.g., via SIMD scoring and more efficient use of memory bandwidth).
Accordingly, we enable \scann quantization in the main experiments and choose corresponding available quantization settings based on the ablation in Table~\ref{tab:scann_ablation}.

\noindent
\uline{Hyperparameter Tuning.}
\textbf{(i) HNSW:}
Our evaluation focuses on the runtime behavior of the search strategies for a fixed HNSW construction across datasets.
All experiments use the same HNSW index built with fixed construction parameters ($M{=}32$, \textit{ef\_construction}{=}200), following prior system settings~\cite{sehgal2025navix}.
For each algorithm, we tune its run-time parameters, primarily \textit{ef\_search} and the (\textit{max\_scan\_tuples} in \pgvector), and use the configuration that yields the highest QPS at 95\% recall.
\textbf{(ii) \scann:}
For filtered \scann, we tune the main run-time knobs that control the candidate budget and the accuracy/throughput trade-off: the number of leaves to scan and the reordering factor (to offset quantization error). We fix construction parameters: \textit{num\_leaves}{=}10K and \textit{max\_num\_levels}{=}1 for OpenAI-5M, and \textit{num\_leaves}{=}100K and \textit{max\_num\_levels}{=}2 for the other datasets.  All results report the average over five independent runs.

\noindent \textbf{Database Configurations.}
To eliminate the possible disk I/O variability, we utilized the \textit{pg\_prewarm} extension to fully load both the table and the index into the database buffer cache before each experimental run. 
The scope of this work focuses on in-memory use cases. By removing disk latency, we demonstrate that even under ideal memory-resident conditions, the architectural overhead of the database engine alone is enough to alter algorithmic trade-offs compared to libraries.
The \postgresql instance we use is set with shared\_buffers = 64GB and maintenance\_work\_mem = 64GB. 

\noindent \textbf{Datasets.}
To ensure a comprehensive and robust evaluation, we selected four datasets that span a wide range of size scales, distance metrics, vector dimensionalities, and intrinsic query difficulties (Table~\ref{tab:dataset_char}). 
Our selection includes datasets with varying costs for distance computations: low (\texttt{\siftten}~\cite{jegou2011Sift}, \texttt{\texttoimage}~\cite{simhadri2022yantti}) and high (\texttt{\cohere}~\cite{Cohere10M}, \texttt{\openai}~\cite{openai5M}), primarily driven by vector dimensionality. 
Furthermore, these datasets exhibit different levels of query hardness, providing a challenging testbed. 
Specifically, the \texttt{\texttoimage} workload includes out-of-distribution (OOD) queries, which are difficult cases in which a query vector may not be close to any data points satisfying the filter. 

\noindent \textbf{Relative Costs.}
We have also calculated the relative cost of distance computations compared to bitmap probing (Table~\ref{tab:dataset_char} Dist-Filt. Rel. Cost column). These results are from running filter checks and distance computations in isolation, without utilizing \postgresql, thereby emulating the relative cost incurred by queries served by a library (\hnswlib).

\noindent \textbf{Workloads.}
For each dataset, we generate a comprehensive query workload designed to test algorithm performance across a spectrum of filter conditions. From a base set of 100 queries, we create variants covering nine distinct filter selectivities, ranging from highly-selective / low-selectivity (0.01) to non-selective / high-selectivity (0.9), and five vector-predicate correlations. 
This methodology yields 4,500 unique queries per dataset ($100 \times 9 \times 5$), enabling fine-grained analysis of how each FVS method behaves across widely varying scenarios.
Each reported data point represents the average performance across the 100 base queries for a given experimental configuration (i.e., correlation and selectivity).

\noindent \textbf{Query Execution in \postgresql.}
In Section~\ref{sec:filter-workload-generation}, we describe how we generate the queries. Each query is described by a query vector and a bitmap of the rowIDs that satisfy a filter with the specified selectivity and correlation. Our experimental setup simulates the execution of a query plan that applies the filters first, produces a bitmap of rowIDs that satisfy the filter, and then performs the vector search.
During the HNSW graph traversal, checking if a vector satisfies the filter becomes a probe into this bitmap. For each query, we report the latency of the FVS algorithm by instrumenting the index access method, recording the wall-clock time from the beginning of the index scan to its completion. This measurement exclusively covers the time spent searching the index, excluding external overheads such as bitmap generation and query planning.

The execution strategy, where the filter is applied first in order to produce a bitmap, is commonly used to avoid switching between vector index traversal and filter evaluation, polluting the caches and the buffer pool ~\cite{sehgal2025navix}. If filters are expensive to apply, evaluating them for the entire input of the vector search might be inefficient. In this case, a query optimizer may evaluate the filters during the traversal of the vector index or choose a post-filtering method. 
In our evaluation, we focus on inline filtering methods that leverage precomputed bitmaps, modeling non-expensive filters.

\noindent \textbf{Query Hardness.} To formally quantify the inherent difficulty of the search task for each dataset, we adopt two key metrics, Local Intrinsic Dimensionality (LID)~\cite{Amsaleg2015} and Local Relative Contrast (LRC)~\cite{he2012lrc}, which have been extensively used in prior work to characterize vector search workloads~\cite{chatzakis2025darth, wang2024steiner, voruganti2025mirage, azizi2025graph, aumuller2021roleoflid, ceccarello2025queryworkloads, wei2025subspace}. LID measures the effective dimensionality in the local neighborhood of a query point, with higher values indicating a more complex search space. LRC assesses the correlation between unfiltered nearest neighbors and true filtered nearest neighbors, where higher values between 0 and 1 suggest a harder filtered search task. Table~\ref{tab:dataset_char} reports these values, confirming the diverse query hardness across our datasets, with \texttt{text2image10M} being particularly challenging.

\noindent
\textbf{Metrics.}
We evaluate system performance using \textbf{Query Per Second (QPS) at 95\% Recall@10}, a standard setting in related literature~\cite{wang2024steiner,chatzakis2025darth}. To dissect sources of latency, we report four metrics: the number of Distance Computations, Filter Checks, Graph Hops (for graphs) and Leaves (for \scann), Reordering (for \scann), and 8KB Page Accesses.

%% file: 9-evaluation-analysis.tex
\section{Evaluations and Analysis}

\subsection{Performance Results}

\begin{figure}[t]
\centering
\includegraphics[width=\textwidth]{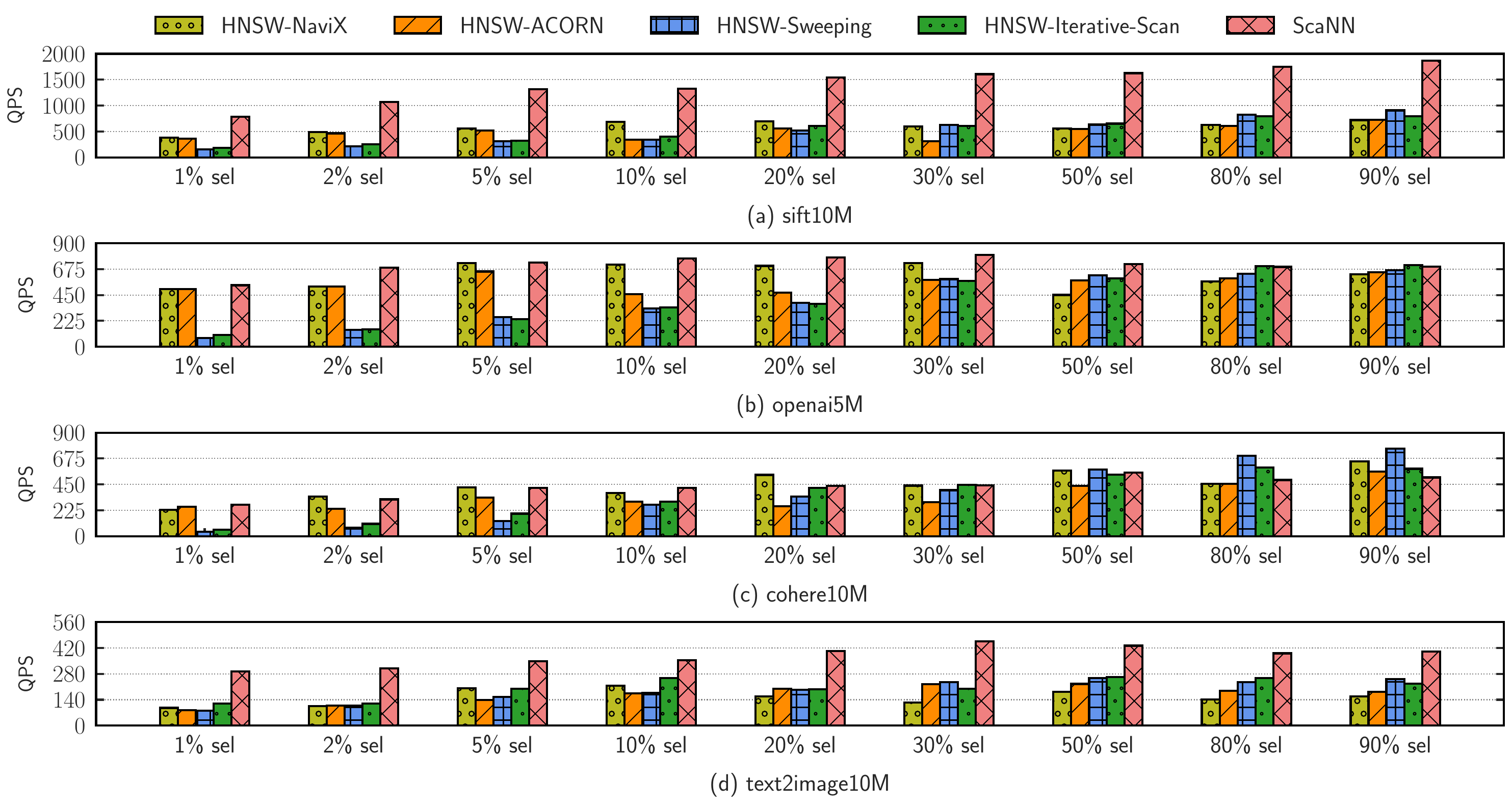}
\caption{
QPS at 95\% Recall@10 for methods across various selectivities on four datasets with no vector-predicate correlation.
}
\label{fig:qps_recall_all_workloads}
\end{figure}

\begin{table}[t]
\centering
\caption{Breakdown of internal index statistics across varying selectivities on OpenAI-5M (no correlation). Colors indicate relative performance within each metric group (green=lower/better, red=higher/worse). Bold indicates the best value per row.}
\label{tab:breakdown_stats}
\resizebox{\textwidth}{!}{%
\setlength{\tabcolsep}{3pt}
\begin{tabular}{lccccccccccccccccc}
\toprule
\multicolumn{1}{c}{} & \multicolumn{4}{c}{\textbf{Distance Computations}} & \multicolumn{4}{c}{\textbf{Filter Checks}} & \multicolumn{4}{c}{\textbf{Hops / (Leaves \& Reordering)}} & \multicolumn{4}{c}{\textbf{Page Access}} \\
\cmidrule(lr){2-5} \cmidrule(lr){6-9} \cmidrule(lr){10-13} \cmidrule(lr){14-17}
\textbf{Sel.} & NaviX & ACORN & Sweep. & ScaNN & NaviX & ACORN & Sweep. & ScaNN & NaviX & ACORN & Sweep. & ScaNN & NaviX & ACORN & Sweep. & ScaNN \\
\midrule
1\% & \cellcolor[RGB]{217,234,211}246 & \cellcolor[RGB]{217,234,211}\textbf{238} & \cellcolor[RGB]{249,224,224}23.0K & \cellcolor[RGB]{217,234,211}1.4K & \cellcolor[RGB]{249,227,224}71.8K & \cellcolor[RGB]{249,224,224}71.6K & \cellcolor[RGB]{249,224,224}\textbf{2.6K} & \cellcolor[RGB]{249,224,224}143.9K & \cellcolor[RGB]{231,239,217}17.8 & \cellcolor[RGB]{217,234,211}\textbf{17.7} & \cellcolor[RGB]{249,224,224}1.1K & \cellcolor[RGB]{249,224,224}150/62 & \cellcolor[RGB]{225,237,214}1.2K & \cellcolor[RGB]{217,234,211}\textbf{1.2K} & \cellcolor[RGB]{249,224,224}24.0K & \cellcolor[RGB]{249,224,224}4.5K \\
2\% & \cellcolor[RGB]{221,235,212}\textbf{357} & \cellcolor[RGB]{226,237,215}408 & \cellcolor[RGB]{244,244,222}9.4K & \cellcolor[RGB]{218,234,211}1.9K & \cellcolor[RGB]{250,230,224}66.6K & \cellcolor[RGB]{250,232,225}59.6K & \cellcolor[RGB]{253,247,226}\textbf{1.4K} & \cellcolor[RGB]{253,244,226}96.0K & \cellcolor[RGB]{238,241,220}\textbf{19.8} & \cellcolor[RGB]{226,237,215}20.9 & \cellcolor[RGB]{242,243,222}398 & \cellcolor[RGB]{253,245,226}100/60 & \cellcolor[RGB]{239,242,220}\textbf{1.5K} & \cellcolor[RGB]{248,246,224}1.6K & \cellcolor[RGB]{244,244,222}9.8K & \cellcolor[RGB]{253,247,226}3.2K \\
5\% & \cellcolor[RGB]{231,239,217}627 & \cellcolor[RGB]{238,241,220}\textbf{614} & \cellcolor[RGB]{229,238,216}5.0K & \cellcolor[RGB]{223,236,213}3.8K & \cellcolor[RGB]{251,246,225}36.2K & \cellcolor[RGB]{252,247,226}34.9K & \cellcolor[RGB]{239,242,220}\textbf{910} & \cellcolor[RGB]{245,244,223}76.9K & \cellcolor[RGB]{217,234,211}\textbf{13.0} & \cellcolor[RGB]{233,240,218}23.0 & \cellcolor[RGB]{228,238,215}197 & \cellcolor[RGB]{245,244,223}80/60 & \cellcolor[RGB]{217,234,211}\textbf{1.1K} & \cellcolor[RGB]{237,241,219}1.4K & \cellcolor[RGB]{229,238,216}5.2K & \cellcolor[RGB]{240,242,221}2.8K \\
10\% & \cellcolor[RGB]{241,243,221}\textbf{886} & \cellcolor[RGB]{252,240,226}1.1K & \cellcolor[RGB]{223,236,213}3.3K & \cellcolor[RGB]{225,237,214}4.8K & \cellcolor[RGB]{239,242,220}24.5K & \cellcolor[RGB]{253,245,226}40.4K & \cellcolor[RGB]{223,236,213}\textbf{359} & \cellcolor[RGB]{226,237,215}48.2K & \cellcolor[RGB]{217,234,211}\textbf{12.8} & \cellcolor[RGB]{229,238,216}21.9 & \cellcolor[RGB]{222,235,213}107 & \cellcolor[RGB]{226,237,215}50/95 & \cellcolor[RGB]{222,236,213}\textbf{1.2K} & \cellcolor[RGB]{249,228,224}1.9K & \cellcolor[RGB]{223,236,213}3.4K & \cellcolor[RGB]{223,236,214}2.2K \\
20\% & \cellcolor[RGB]{252,247,226}\textbf{1.2K} & \cellcolor[RGB]{251,236,225}1.2K & \cellcolor[RGB]{219,235,212}2.2K & \cellcolor[RGB]{237,241,219}9.7K & \cellcolor[RGB]{227,238,215}12.6K & \cellcolor[RGB]{230,239,216}14.6K & \cellcolor[RGB]{219,234,212}\textbf{219} & \cellcolor[RGB]{226,237,215}48.2K & \cellcolor[RGB]{218,234,211}\textbf{13.2} & \cellcolor[RGB]{252,242,226}32.3 & \cellcolor[RGB]{219,234,211}64.3 & \cellcolor[RGB]{226,237,214}50/80 & \cellcolor[RGB]{232,239,217}\textbf{1.4K} & \cellcolor[RGB]{251,238,225}1.8K & \cellcolor[RGB]{219,235,212}2.3K & \cellcolor[RGB]{221,235,213}2.1K \\
30\% & \cellcolor[RGB]{253,247,226}\textbf{1.2K} & \cellcolor[RGB]{249,226,224}1.5K & \cellcolor[RGB]{218,234,211}1.7K & \cellcolor[RGB]{249,246,225}14.5K & \cellcolor[RGB]{222,235,213}6.8K & \cellcolor[RGB]{225,237,214}9.8K & \cellcolor[RGB]{218,234,211}\textbf{178} & \cellcolor[RGB]{226,237,215}48.2K & \cellcolor[RGB]{217,234,211}\textbf{12.8} & \cellcolor[RGB]{249,224,224}41.7 & \cellcolor[RGB]{217,234,211}48.3 & \cellcolor[RGB]{226,237,214}50/80 & \cellcolor[RGB]{231,239,217}\textbf{1.4K} & \cellcolor[RGB]{249,224,224}2.0K & \cellcolor[RGB]{218,234,211}1.8K & \cellcolor[RGB]{221,235,213}2.1K \\
50\% & \cellcolor[RGB]{249,224,224}2.2K & \cellcolor[RGB]{249,224,224}1.6K & \cellcolor[RGB]{217,234,211}\textbf{1.5K} & \cellcolor[RGB]{251,235,225}24.1K & \cellcolor[RGB]{222,236,213}7.4K & \cellcolor[RGB]{220,235,212}4.7K & \cellcolor[RGB]{217,234,211}\textbf{154} & \cellcolor[RGB]{226,237,215}48.2K & \cellcolor[RGB]{250,246,225}\textbf{24.0} & \cellcolor[RGB]{250,232,225}37.4 & \cellcolor[RGB]{217,234,211}39.9 & \cellcolor[RGB]{226,237,214}50/85 & \cellcolor[RGB]{249,224,224}2.4K & \cellcolor[RGB]{250,233,225}1.8K & \cellcolor[RGB]{217,234,211}\textbf{1.5K} & \cellcolor[RGB]{222,236,213}2.2K \\
80\% & \cellcolor[RGB]{252,239,225}1.5K & \cellcolor[RGB]{249,227,224}1.5K & \cellcolor[RGB]{217,234,211}\textbf{1.4K} & \cellcolor[RGB]{249,224,224}30.9K & \cellcolor[RGB]{217,234,211}2.5K & \cellcolor[RGB]{217,234,211}2.1K & \cellcolor[RGB]{217,234,211}\textbf{149} & \cellcolor[RGB]{220,235,212}38.6K & \cellcolor[RGB]{249,224,224}37.1 & \cellcolor[RGB]{253,243,226}\textbf{32.1} & \cellcolor[RGB]{217,234,211}38.2 & \cellcolor[RGB]{219,235,212}40/100 & \cellcolor[RGB]{248,245,224}1.6K & \cellcolor[RGB]{249,246,225}1.6K & \cellcolor[RGB]{217,234,211}\textbf{1.4K} & \cellcolor[RGB]{217,234,211}2.0K \\
90\% & \cellcolor[RGB]{253,243,226}1.4K & \cellcolor[RGB]{250,231,224}1.4K & \cellcolor[RGB]{217,234,211}\textbf{1.3K} & \cellcolor[RGB]{249,224,224}30.5K & \cellcolor[RGB]{217,234,211}1.7K & \cellcolor[RGB]{217,234,211}1.6K & \cellcolor[RGB]{217,234,211}\textbf{138} & \cellcolor[RGB]{217,234,211}33.9K & \cellcolor[RGB]{249,224,224}37.5 & \cellcolor[RGB]{252,243,226}\textbf{32.1} & \cellcolor[RGB]{217,234,211}35.3 & \cellcolor[RGB]{217,234,211}35/120 & \cellcolor[RGB]{236,241,219}1.4K & \cellcolor[RGB]{236,241,219}1.4K & \cellcolor[RGB]{217,234,211}\textbf{1.4K} & \cellcolor[RGB]{217,234,211}2.0K \\
\bottomrule
\end{tabular}%
}
\end{table}

\begin{figure}[]
\centering
\includegraphics[width=0.595\linewidth]{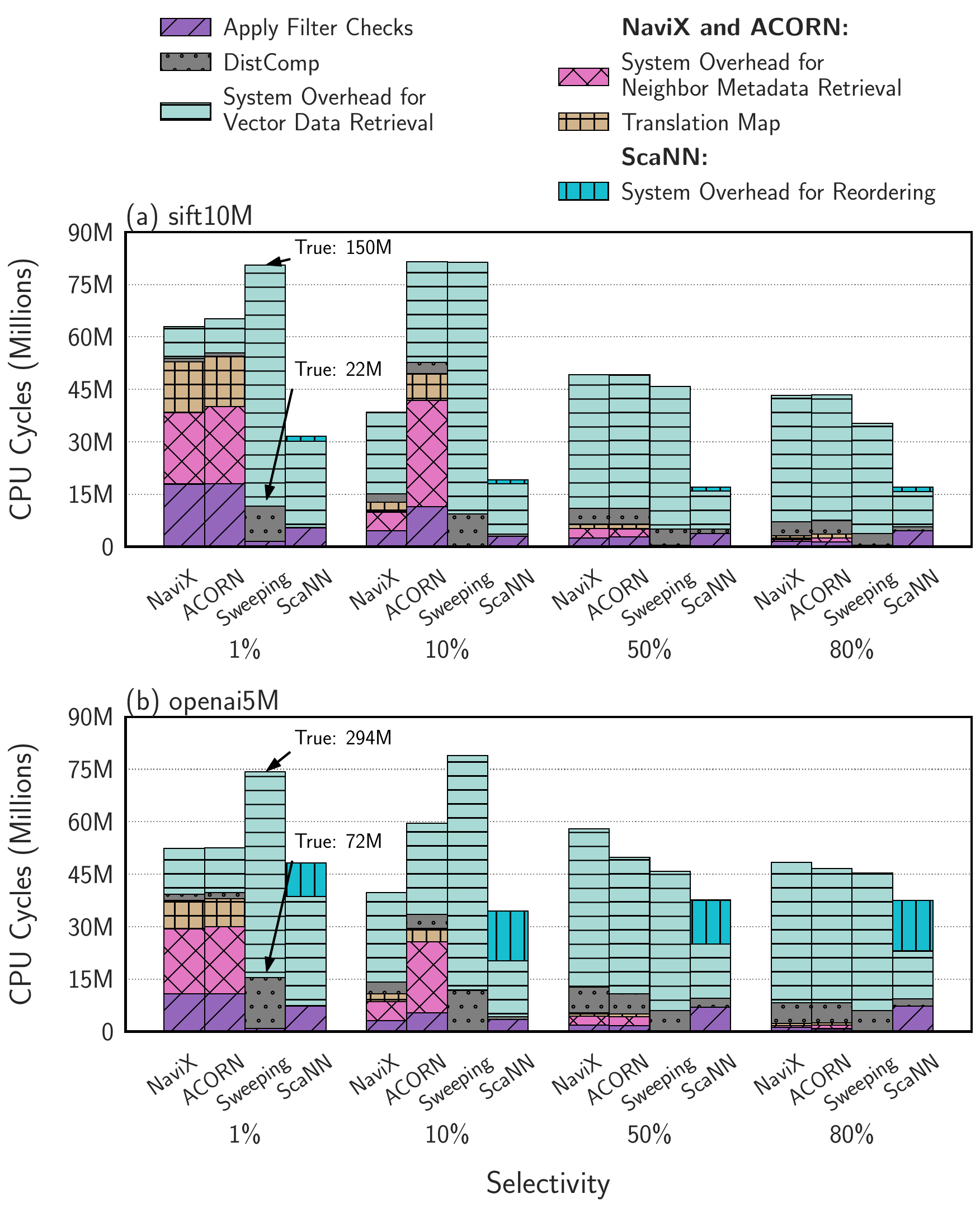}
\caption{ 
Latency breakdown (CPU cycles) of a query running for 1, 10, 50, and 80\% selectivities on the \oai dataset without correlation. Bars (1\% sel. for \sweeping) are clipped at the plot’s y-axis limit for readability; annotations “True: X” report the full (unclipped) cycle count.
}
\label{fig:cycle-breakdown-acorn-sweeping}
\end{figure}

\textbf{Trends across datasets--}
Figure~\ref{fig:qps_recall_all_workloads} shows QPS performance for filter selectivities ranging from 1\% to 90\% (without correlation) across the four datasets described in Section~\ref{sec:exp_setup}. 
We see a few common trends across the different datasets:

\begin{enumerate} [wide=0pt, nosep]
    \item Trend 1: The clustering-based approach (\scann) performs in general better (2-3x) than the graph approaches on low-dimensional vectors (\siftten and \texttoimage). The difference is smaller for high-dimensional vectors (openai5M and cohere10M).
    \item Trend 2: Filter-first methods (\navix and \acorn) perform better for low selectivities compared to traversal-first methods (\sweeping and \iterativescan), which perform better for larger selectivities.
    
    \item Trend 3: The \navix-Directed heuristic consistently outperforms \acorn at low-to-mid selectivities (5\%–30\%).
\end{enumerate}

\noindent These trends highlight that optimal performance is driven by the interplay between algorithmic efficiency and the system-level overheads outlined in Table~\ref{tbl:indexes}. 
First, regarding Trend 1, \scann's superiority on low-dimensional vectors is explained by its sequential access pattern, which mitigates the \emph{random memory access amplification} inherent in graph traversals. As dimensionality increases, for \scann, the number of vectors fitting on a page drops drastically (e.g., by 10$\times$), which seriously affects the performance for the rescoring part of the ScaNN method that uses the full original dimensions and non-quantized values. Hence, the random access penalty of the graph is less amplified relative to the \scann, as even sequential approaches must fetch significantly more pages to retrieve the same number of vectors, causing performance to converge.

Second, regarding Trend 2, the results align with the motivational findings from Figure~\ref{fig:motivation-plot}. We observe a distinct crossover point where the overhead of filter-first exploration outweighs its benefits. Filter-first methods, such as \acorn, prioritize filtering neighbors before scoring them to reduce distance computation costs; however, as selectivity increases, valid neighbors become abundant. At this point, the cost of maintaining the predicate subgraph exceeds the savings, thereby making traversal-first methods more efficient. 
Within low to mid selectivities where \navix-Directed heuristic is activated, it outperforms \acorn because the Directed heuristic effectively guides the search towards valid regions at fewer hops (Table ~\ref{tab:breakdown_stats}: 5\%-30\%) than \acorn requires (Trend 3).

\subsection{Where Does the Time Go?}
\label{subsec:where-time-goes}

To understand the performance characteristics of different FVS methods, we perform a detailed breakdown of query execution time (Figure~\ref{fig:cycle-breakdown-acorn-sweeping}) and internal search metrics (Table~\ref{tab:breakdown_stats}).

\subsubsection{Search metrics.}
\label{subsub:filt-dist-count}

\uline{(i) For graph-based methods}, filter-first approaches (\navix/\acorn) perform fewer hops than traversal-first (\sweeping/\iterativescan) to reach the same recall, since \navix/\acorn can filter and score two-hop neighbors without first hopping to their immediate neighbors (two-hop neighbor expansion).
At low selectivity (1\%), they perform significantly fewer distance computations (\navix: 246, \acorn: 238) compared to \sweeping (23.0K), achieving up to 100$\times$ reduction through predicate-subgraph navigation.
However, this reduction comes at the cost of more filter checks: \navix and \acorn evaluate 71.8K and 71.6K filters, respectively, compared to \sweeping's 2.6K, as graph traversal opportunistically checks many candidates during navigation.
Page access patterns provide an additional direct explanation for performance differences: at 1\% selectivity, \sweeping incurs 24.0K page accesses, while \navix and \acorn maintain only 1.2K through selective navigation.
\uline{(ii) For \scann,} it exhibits different behavior due to its cluster-based architecture.
At low selectivity, it incurs significant number of filter checks (143.9K at 1\%) because it evaluates filters for every page of every opened leaf.
As selectivity increases, \scann requires fewer leaves to reach the target recall, so filter checks decrease while distance computations increase.
For page access, \scann achieves 4.5K at 1\% selectivity and improves to \textasciitilde2K at mid selectivities (10--30\%), but still larger than the optimal graph-based method (\navix), at this selectivity range.
These internal metrics explain only part of the story. 
Hops between filter-first and traversal-first methods do not incur the same per-step cost, and metrics are not uniformly better for one method across all selectivities. Moreover, page accesses across methods do not follow the same sequence of steps, either.
\emph{This makes these metrics not fully comparable across methods, motivating a more in-depth study of where time is actually spent, which we present next.}

\subsubsection{Where the Time Really Goes.}
\label{sec:profiling}

Figure~\ref{fig:cycle-breakdown-acorn-sweeping} shows the time breakdown for each of the search methods (\iterativescan exhibits similar performance to \sweeping, therefore we don't show its breakdown). Each color-coded component in the legend corresponds to a numbered operational step illustrated in Figures~\ref{fig:acorn-pgvector} and~\ref{fig:scann-pgvector}.

As described in Section~\ref{subsec:search-steps}, each search algorithm is composed of several actions.
For \textbf{graph-based methods}, these range from operations like accessing an index page to retrieving neighbor's indexTIDs (step \circled{1} for all graph methods) and accessing a direct neighbor's index page to acquire 2-hop neighbors (step \circled{2} in filter-first), to accessing the Translation Map (step \circled{3} in filter-first), applying filter checks (step \circled{4} in filter-first), and more complex multi-stage process of distance computation and element materialization (step \circled{5} in filter-first).
For \textbf{\scann}, the operational flow is different.
A leaf may logically contain multiple pages. 
For each page, \scann first retrieves all heapTIDs for probing bitmaps (step \circled{1} for ScaNN), then executes relatively simpler steps: filtering (step \circled{2} for ScaNN) and distance computation (step \circled{3} for ScaNN).
Unlike graph methods, \scann does not require complex traversal steps such as neighbor storage/lookups and maintaining various navigation lists.
Our profiling reveals that \emph{the vast majority of query time for all algorithms is spent on low-level system overheads such as page access and data handling operations}.
These costs are not redundant; they are a direct consequence of maintaining the transactional consistency and isolation guarantees of an ACID-compliant database.
This presents a fundamental challenge when implementing high-throughput search algorithms within a database storage manager.

Figure~\ref{fig:cycle-breakdown-acorn-sweeping} shows that at low selectivities, filter-first approaches divide their latency among many more components than traversal-first methods.
For example, \navix and \acorn's latency at 1\% selectivity is distributed across Translation Map access, neighbor metadata retrieval, filter checks, and distance computations.
In contrast, \sweeping's performance is mainly dominated by system overhead for vector data retrieval rather than distance computation itself --- at 1\% selectivity, OpenAI-5M, vector retrieval consumes over 300M CPU cycles (shown truncated with ``True: 300M'' annotation), far exceeding the actual computational work.
These results signify that although \navix/\acorn's overall performance is preferred at low selectivities, it is affected by many system-level factors not present in the \hnswlib library.
The proportion by which these factors affect overall performance correlates with filter selectivity, explaining the shift in crossover point between the \hnswlib and \pgvector results (Figure~\ref{fig:motivation-plot}).
For \textbf{\scann}, one distinct component is the reordering step that requires (i) retrieval of full-precision vectors from the heap (system overhead for reordering) and (ii) sorting by distance to identify the top-k candidates.
This is especially significant for OpenAI-5M, where quantization compresses vectors from 1536$\times$4 bytes to approximately 193$\times$1 byte ($\approx$32$\times$ compression), requiring nearly one page access (8196 / 1536$\times$4 $\approx$1) per vector during reordering.

\subsubsection{System Implications: Graph v.s. Tree}
\label{sec:system-implications}
Taking metrics and profiling results together, we observe a phenomenon: \scann incurs more filter checks and page accesses at low selectivities, and more distance computations at high selectivities, compared to graph-based methods, yet it often consumes fewer CPU cycles overall.
This paradox can be explained by considering the fundamental differences between graph-based and tree-based index structures in a page-based storage system.
\uline{(i) Memory access patterns: random vs. sequential.}
Graph methods (HNSW) exhibit \textbf{random access patterns} during traversal.
Each hop to a neighbor node requires loading different pages distributed across the buffer pool, resulting in poor spatial locality and preventing effective hardware prefetching.
In contrast, \scann's tree-based clustering enables \textbf{sequential access within partitions}.
Once a cluster leaf is selected, \scann scans vectors within that partition sequentially across buffer pool pages, benefiting from hardware prefetching and cache line fills.
This sequential pattern reduces memory latency despite touching more data.
\uline{(ii) The ``system tax'' for complex heuristics.}
Graph methods impose a \textbf{system tax} for their sophisticated navigation heuristics.
The Translation Map overhead (5--15M cycles) and neighbor metadata retrieval (15--25M cycles) represent fixed costs of maintaining the HNSW structure at low-mid selectivities. 
\uline{(iii) Hardware utilization.}
\scann's sequential vector scanning is \textbf{highly amenable to SIMD vectorization}.
Filter evaluation on cluster members enables batched bitmap probing, and distance computations within a cluster can leverage SIMD instructions for parallel processing.
The uniform access pattern (load cluster pages $\rightarrow$ filter $\rightarrow$ compute distances) maps naturally to modern CPU vector units.
In contrast, graph methods exhibit poor hardware utilization due to irregular access patterns (at page granularity).
Each graph hop involves a conditional decision (which neighbor to visit next?), making it difficult to batch operations.
Data-dependent control flow further limits vectorization opportunities.
This explains why graph methods exhibit higher end-to-end latencies despite fewer distance computations: the work cannot be effectively parallelized.
\uline{(iv) Index footprint and cache locality.}
Graph methods require storing the full HNSW structure (vertices and neighbor lists), resulting in substantial memory overhead beyond the raw vectors.
The random access pattern means that graph metadata frequently evicts useful data from cache.
This is particularly problematic for large-scale indexes where the graph structure cannot fit in the L3 cache, forcing frequent main memory accesses.

\noindent
\textbf{Takeaway.}
The graph vs. tree tradeoff reveals that \emph{system-level efficiency can offset algorithmic efficiency}.
For instance, filter-first methods can achieve fewer distance computations and lower page access --- yet \scann's sequential access patterns, SIMD-friendly operations, and cache locality, yield competitive overall throughput.
Dataset characteristics (e.g., vector dimensionality) can tip the balance, allowing graph methods to bridge the gap in certain workloads.
Therefore, no single structure dominates across all scenarios.
\emph{This highlights that any correlation between lightweight libraries and \postgresql system performance is, at best, coincidental. We encourage the community to be aware of system-level implications when proposing new methods targeting vector search within database systems.}



\begin{figure}[t]
\centering
\includegraphics[width=0.635\linewidth]{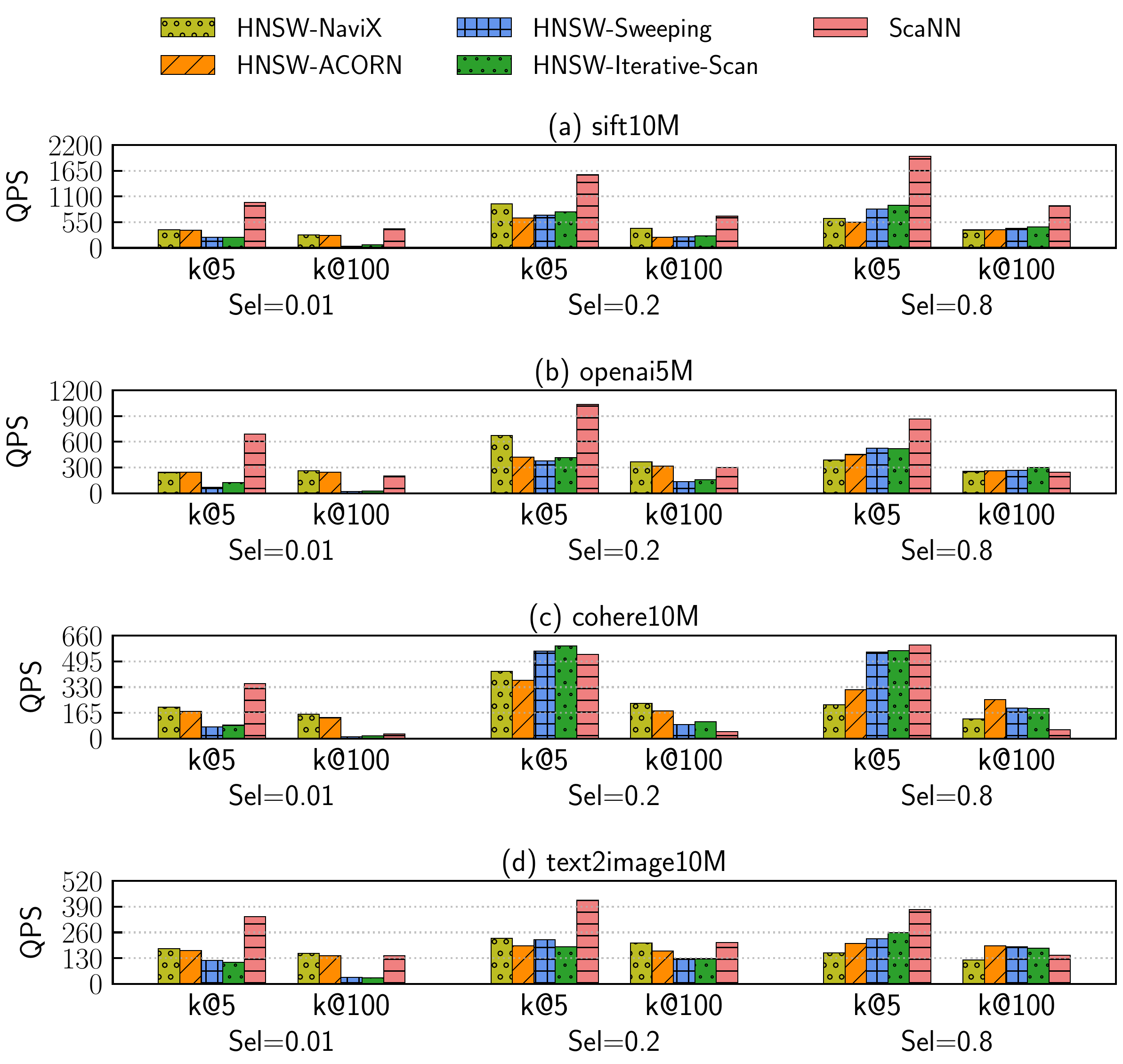}
\caption{
Varying LIMIT $k$ on all datasets (no correlation).
}
\label{fig:various_k}
\end{figure}

\subsection{Sensitivity to Result Size (LIMIT $k$)}
\label{sec:vary-k}
We evaluate $k\in\{5,100\}$ at 95\% Recall@$k$ (Figure~\ref{fig:various_k}).
When the requested result size increases, the query returns more \emph{filter-valid} neighbors and makes the search procedure perform more work --- either by traversing more of the graph (graph methods) or by scanning more leaves (\scann).

This growth is modest for filter-first methods (\navix/\acorn), which traverse the predicate-induced subgraph, but substantial for traversal-first baselines. At 1\% selectivity, OpenAI-5M, for example, \navix’s hop count is changed reasonably when moving from $k{=}5$ to $k{=}100$ (12.0$\rightarrow$22.4, +86\%), while \sweeping increases sharply (991.6$\rightarrow$6307.6, +536\%). In terms of graph hops, while \navix grows by +106\%, \sweeping grows by +599\%. \scann similarly scales up its effort with $k$ (leaves scanned: 122$\rightarrow$391, +220\% at 1\% selectivity; +294\% on average), which narrows its advantage in the high-$k$, low-selectivity setting.




\subsection{Non-Monotonic Performance}

An observation from our results (Figure ~\ref{fig:qps_recall_all_workloads}) is that the performance of adaptive indices (\navix/\acorn) is not strictly monotonic with respect to selectivity.
This phenomenon is explained by the interaction between \emph{adaptive heuristics} and the different \emph{system overheads} of each of these techniques. As detailed in Table~\ref{tab:breakdown_stats}, the search passes through distinct phases: \textbf{i) Aggressive filtering (1--5\%):} Two-hop neighbors \textit{Filter Checks} minimize \textit{Distance Computations}. \textbf{ii) Transition to Directed heuristic (\navix) / Skip some 2-hop (\acorn) (5--30\%):} As selectivity increases, the graph exploration expands (increasing distance computations), but the heuristic has not yet fully switched to a less filter-focused search strategy, causing filter checks to remain high.
\textbf{Converge to normal graph-traversal ($>=$50\%):} The algorithm detects sufficient density and switches to skipping checks or standard 1-hop traversal. Filter checks drop sharply, and the bottleneck shifts entirely to distance computations and system overheads.
Figure~\ref{fig:cycle-breakdown-acorn-sweeping} visualizes these opposing forces: as selectivity increases, filter-related costs shrink, while scoring and system-related costs expand. 
The same observation can be found in the original \navix ~\cite{sehgal2025navix} across selectivities.

For \scann, we observe a slight QPS drop after 50\% selectivity because we execute inline-filtering throughout the search. As selectivity increases, the bitmap sizes grow significantly and can no longer fit in cache easily; thus, we observe increasing overhead for filtering (Figure~\ref{fig:cycle-breakdown-acorn-sweeping}) even as the number of filter checks decreases. 



\subsection{Correlation Effects}

\begin{figure}[]
\centering
\includegraphics[width=0.55\linewidth]{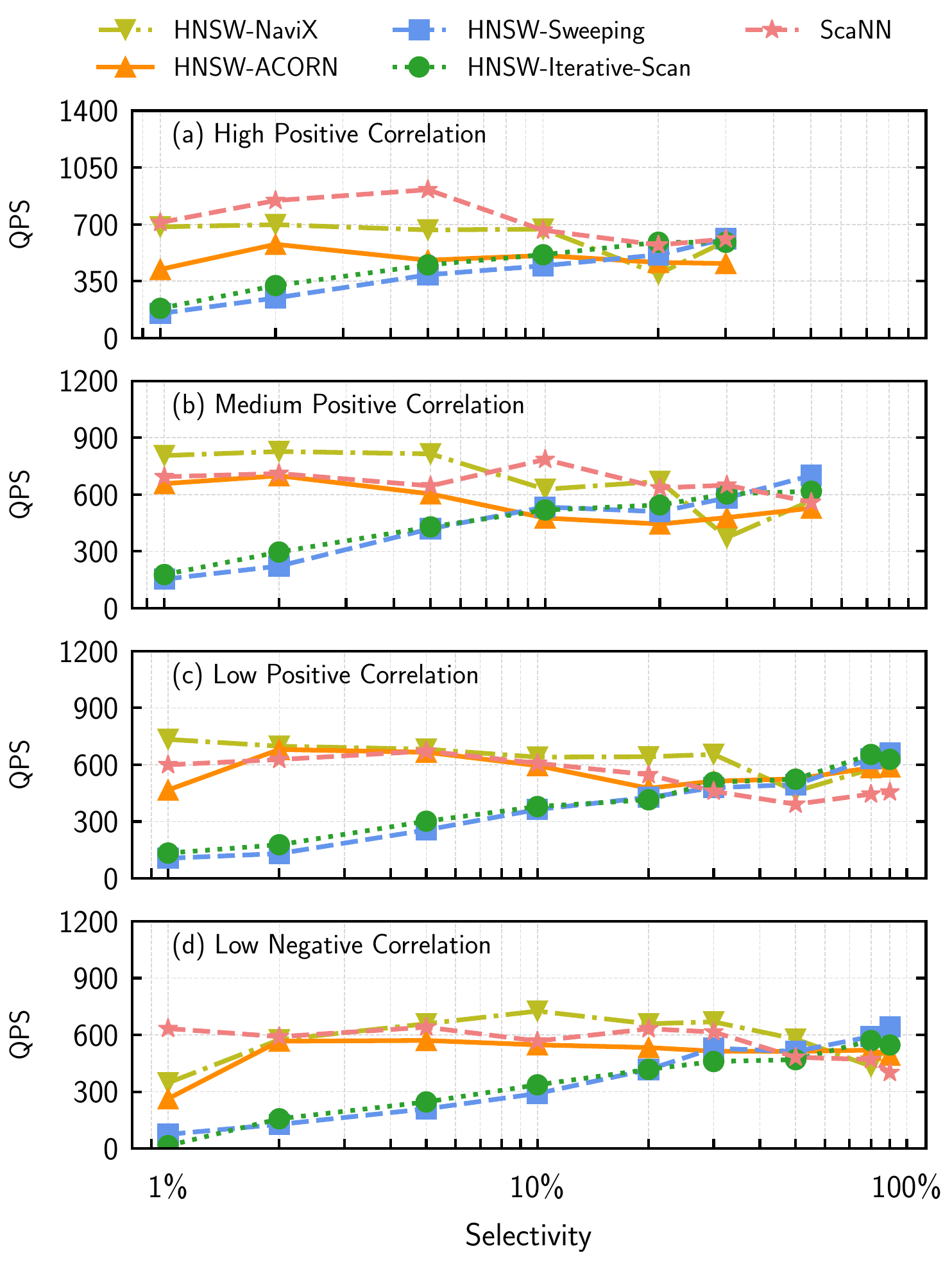}
\caption{
QPS v.s. Selectivity on \oai, showing how vector-predicate correlation alters the performance and crossover point of each FVS algorithm.
}
\label{fig:correlations-effect}
\end{figure}

Vector-predicate correlation is a query property that significantly alters the search effort needed to achieve the target recall. Figure~\ref{fig:correlations-effect} illustrates the performance of five methods on the \oai dataset across selectivities for four different correlation settings. For positive correlation workloads (Figure~\ref{fig:correlations-effect} (a)--(c)), where the nearest neighbors to a query vector are more likely to satisfy the predicate, \navix consistently outperforms \acorn by 1.2--1.7$\times$, while traversal-first methods like \sweeping and \iterativescan improve significantly as selectivity increases, becoming competitive with \navix at selectivities as low as 10\%. \scann achieves strong throughput at low selectivities (711--913 QPS under high positive correlation) but degrades at higher selectivities where its filter-then-rank approach incurs overhead from processing more candidates.

Negative correlations in workloads reveal fundamental differences among algorithm architectures. As shown in Figure~\ref{fig:correlations-effect}(d), when nearest neighbors fail the predicate, graph-based performance degrades at 1\% selectivity: \navix drops by 53\%, \acorn by 44\%, and \iterativescan by 89\% relative to the positive correlation. These methods waste resources exploring graph regions where vectors are close to the query but do not satisfy the predicate. In contrast, \scann demonstrates remarkable robustness—its throughput increases by 6\% under negative correlation, making it the optimal choice for this setting. Because \scann's tree-based partitioning does not rely on graph proximity, it avoids the exploration penalty that graph traversal methods incur. \navix shows a dip around 50--80\% selectivity, where its adaptive heuristic may skip beneficial 2-hop expansion (at 50\%) mostly, and there is not enough connectivity for solely looking at one-hop neighbors; at 90\%, abundant qualifying 1-hop neighbors make the heuristic efficient again.

Despite its sensitivity to negative correlation at very low selectivity, \navix recovers to become the best-performing method from 5--50\% selectivity even under negative correlation. These results confirm that more adaptive graph-based \navix provides superior performance across most settings.



\begin{figure}[t]
\centering
\includegraphics[width=0.695\linewidth]{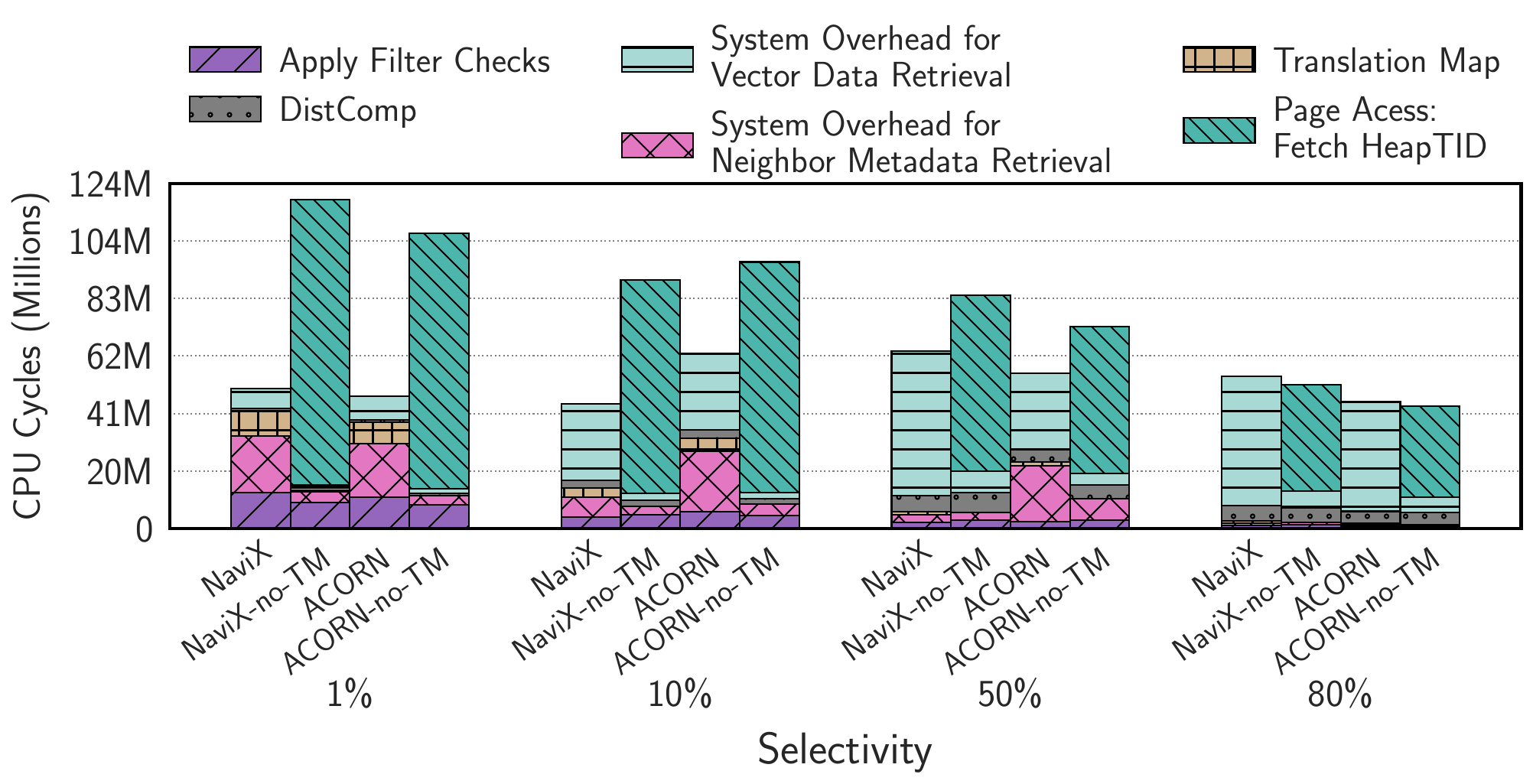}
\caption{
Latency breakdowns with/without a translation map on the \oai dataset (no correlation).
}
\todo{}
\label{fig:ablation-translation-map}
\end{figure}

\begin{table}[t]
\centering
\caption{Comparison of average CPU cycles breakdown between single- v.s. multiple- threads (1T v.s. 16T) at 10\% selectivity on OpenAI-5M with 95\% recall@10 (no correlation).}
\label{tab:thread_profiling}
\small
\setlength{\tabcolsep}{3pt}
\renewcommand{\arraystretch}{1.0}
\begin{tabular}{llcccc}
\toprule
\textbf{Method} & \textbf{Threads} & \textbf{Total Cycles} & \textbf{SysOH\%} & \textbf{DistComp\%} & \textbf{Filter\%}\\
\midrule
NaviX & 1T  & 24.1M         & 55.9\% & 13.0\% & 12.3\% \\
NaviX & 16T & 35.6M (+48\%) & 73.5\% & 7.5\%  & 9.5\%  \\
\midrule
Sweep. & 1T  & 45.0M         & 81.8\% & 18.6\% & 0.2\% \\
Sweep. & 16T & 75.7M (+68\%) & 91.0\% & 11.9\% & 0.2\% \\
\midrule
ScaNN & 1T  & 20.7M         & 84.4\% & 3.1\% & 12.5\% \\
ScaNN & 16T & 32.9M (+59\%) & 86.9\% & 2.5\% & 10.5\% \\
\bottomrule
\multicolumn{6}{l}{\footnotesize{Note: For NaviX, SysOH includes both Vector Data Retrieval and Neighbor Metadata}} \\
\multicolumn{6}{l}{\footnotesize{Retrieval. For ScaNN, SysOH includes both Vector Data Retrieval and Reordering}} \\
\multicolumn{6}{l}{\footnotesize{(Retrieval Vector Data from Heap). Translation map overhead (8-17\%) is omitted}} \\
\multicolumn{6}{l}{\footnotesize{from the table for clarity.}}
\end{tabular}
\end{table}

\subsection{Sources of System Overhead}
\label{subsub:source-system-overhead}

We analyze two DBMS-specific effects behind the “system tax”: ID indirection (via the translation-map ablation) and concurrency.

\subsubsection{Ablation breakdowns of the translation map.}
\label{subsub:ablation-breakdowns}

Figure~\ref{fig:ablation-translation-map} breaks down CPU cycles for \navix and \acorn with and without the translation map (TM).
When TM is disabled, fetching \texttt{heaptid} dominates: across 1\%--50\% selectivity, it accounts for roughly \textbf{60\%--75\%} of total cycles. This overhead stems from the DBMS-specific ID indirection in the system (Table~\ref{tbl:indexes}): 2-hop filtering needs the candidate’s heap tuple identifier (\texttt{heaptid}), which is stored on the index page and thus requires page accesses to resolve. Enabling TM replaces these repeated page accesses with an in-memory lookup.
We also observe that, with TM enabled, the relative shares of other components (e.g., vectors/neighbors retrieval) become more visible. This is primarily a cost-shift effect: disabling TM may warm the relevant index pages and benefit subsequent steps via temporal locality, but the metadata-fetch page accesses still dominate end-to-end runtime. Once TM removes this dominant component, the remaining inherent work becomes the new bottleneck.
At 80\% selectivity, the advantage of TM narrows because the search increasingly relies on 1-hop neighbors and performs substantially fewer 2-hop expansions, reducing the frequency of \texttt{heaptid} resolutions.

\subsubsection{Isolating contention effects}
\label{subsub:contention-analysis}
Our earlier results used 16 concurrent threads. To understand whether contention dominates system overheads and how much concurrent execution (e.g., lock conflicts and cache interference) amplifies these costs, we compare single-thread vs. multi-thread performance.
Table ~\ref{tab:thread_profiling} demonstrates the statistics for one configuration (10\% selectivity, \oai, 95\%recall@10, no correlation) that exhibits representative behavior. System overheads (buffer reads, tuple access), excluding productive computation (distance calculations, bitmap probings), dominate performance even in single-threaded execution: \navix (55.9\%), \sweeping (81.8\%), \scann (84.4\%) of total CPU cycles. This single-thread baseline establishes that high overhead is inherent to \emph{algorithms' intrinsic architectural costs}, independently of the concurrency contention.
Scaling to 16 threads indeed increases total CPU cycles per query (\navix: +48\%, \scann: +59\%, \sweeping: +68\%) as cache sharing and buffer management introduce additional costs.
These results highlight that concurrency amplifies the system overhead costs differently for each method, indicating that some methods may benefit from more system optimizations than others.

%% file: 10-lessons.tex
\section{Observations and Insights}

\noindent
\textbf{\uline{1. System Implications Invalidate the Assumptions.}}
Performance in \postgresql is governed by \emph{page-level data retrieval and metadata management} (buffer lookups, TID indirection) rather than distance computation alone. This shift can change library-observed “viability zones” for algorithms: techniques that appear attractive in libraries, because metadata access is a single pointer dereference, can become \emph{memory-bound} in a DBMS, where every neighbor dereference triggers multiple page accesses. Filter-first traversals may translate into higher \emph{system work per visited node}, which can erode or even nullify their algorithmic advantages, making their end-to-end gains in PG negligible without explicit metadata decoupling. Recent work has begun to adopt this decoupling perspective in \postgresql ~\cite{liu2026postgresqlv}.

\noindent
\textbf{\uline{2. No Clear Winner: Graph vs. Tree.}}
In systems, neither graph nor clustering indexes dominate universally. 
Our experiments show: 
(i) \textbf{Dimensionality}: High-dimensional vectors reduce \scann's sequential-scan advantage, narrowing (or eliminating) the gap with HNSW-style graphs. 
(ii) \textbf{Resilience to $k$}: Graph-based filter-first methods are more robust as $k$ increases (Figure~\ref{fig:various_k}); while \scann must visit more leaves, filter-first methods efficiently identify high-quality qualifying candidates on the predicate subgraph.

\noindent
\textbf{\uline{3. Fewer Hops as the Core Design of Graph.}}
Reducing hops remains the core advantage of graph-based filter-first designs: they can ``tunnel'' through the predicate-induced subgraph to quickly reach filtered candidates. However, this only translates to end-to-end wins if the hop-reduction heuristic does \emph{not} introduce prohibitive per-hop overhead (e.g., extra metadata lookups). Guided expansion (e.g., NaviX-Directed) is the key lesson: it preserves the “fewer hops” benefit, thus reducing wasted exploration.

\noindent
\textbf{\uline{4. A Call for System-Aware Algorithm Design.}}
Our study proposed that many FVS ideas that look strong in library settings can degrade in production DBMSs unless they explicitly account for system overheads. The actionable direction is not “new optimizations in isolation,” but \emph{co-design}: algorithms that (i) expose and minimize system taxes (metadata layout and page access patterns), and (ii) adapt online across selectivity/correlation and query hardness. 


\label{sec:lessons}

%% file: 11-related-work.tex
\section{Related Work}
\label{sec:related_work}


\noindent \textbf{Algorithms for Filtered Vector Search}.
There have been multiple efforts to integrate filters on graph indexes~\cite{gollapudi2023filteredvamana, wang2023nhq, patel2024acorn, weviate, ait2025rwalks, Yin2025DEGEH, Cai2024Navigating, Xu2024iRangeGraph, Zhang2025Efficient, Jiang2025DIGRA, li2025sieve, Liang2025UNIFY, sweeping-wea}. 
Such approaches could be classified into filter-aware and filter-agnostic. 
The former integrates the filter attributes as part of the index building process, evaluating filter information to decide which connections to build, given the common filters among points~\cite{gollapudi2023filteredvamana} or by modifying the distance metric~\cite{wang2023nhq} to incorporate filter similarity. 
The latter family of mechanisms (also called inline filtering methods) is unaware of the filter attribute columns and operations at build time~\cite{patel2024acorn, sehgal2025navix, sweeping-wea}.
While they still utilize an index for searching, they combine applying the filters as they traverse the index.
While pre- and post-filtering approaches are technically also filter-agnostic, they remain complementary to the inline set of solutions. 
In this work, we focus on filter-agnostic mechanisms since we are interested in serving a wide variety of use cases that cannot assume the filter attribute columns or operations are known or can be used at build time. 
See ~\cite{fvs-survey} for more details on the diversity of filtered vector search methods.

\noindent \textbf{Vector Data Management Systems}.
The growing interest in the applications of vector search in recent years has led to an increasing need for systematic approaches in vector data processing.
Such approaches are either added as additional functionalities on top of classic database systems~\cite{scann4alloydb, oracle-ai-vector-search, pgvector, upreti2025cosmosdb, zhang2024fundamentallimitspg}, or have led to the emergence of standalone, vector-native database systems~\cite{pinecone-vector-database, weviate, wang2021milvus}.
In fact, some works already study the overheads of such general systems when adding vector search features~\cite{zhang2024fundamentallimitspg}, although they study vector search in isolation, without filters.
Production-level database systems that support vector search operations enable vectors to coexist in tables alongside structural data, which can serve as filters in Filtered Vector Search processes.
In such systems, filter-agnostic vector search algorithms provide the most viable solutions, as the continuous addition of new data leaves no room for assumptions about the data that might serve as filters.

\noindent \textbf{Filtered Vector Search Evaluations}.
Current works evaluate different kinds of filtered vector search algorithms using them as standalone libraries, running in isolation from other system-related aspects that might greatly affect their performance in a real-world production-level database system~\cite{Zhan2025BigVectorBench, li2025attribute, sehgal2025navix, patel2024acorn, ait2025rwalks}.
At the same time, public information about the performance of those algorithms in production-level database systems usually provides limited results and insights~\cite{weviate, milvus-blog-benchmark, aws-aurora-blog-benchmark, pinecone-blog-benchmark}.
To the best of our knowledge, our work unleashes novel critical performance insights in a complex and general database system, with a breakdown of the manifold sources of performance cost.
In addition, our evaluation is the first one to deep-dive into various costs in a real system, together with the implications of filter selectivities and correlations, all combined, which determines the optimal search algorithm for each case. 

%% file: 13-conclusion.tex
\section{Future Work and Conclusions}
Our benchmark evaluates filter-agnostic FVS algorithms across diverse selectivities and correlations within a \postgresql database system. We show that the system-level overheads create important performance trade-offs. These findings provide a cost-based framework for practitioners and highlight the necessity for future research to shift from isolated library evaluations to the complexities of system-integrated algorithm design. 
There is a clear need for coordinated innovation in algorithms and databases to deliver the best of both worlds: matching the performance of best-in-class in-memory vector search algorithms when RAM is plentiful, while leveraging the transactional consistency and reliability of a database for out-of-core datasets.
A promising approach in this direction is the use of columnar engines to host both the index as well as the columns that participate in filtered vector search. Columnar engines have shown their ability to drastically reduce the page access overhead and other system overheads while maintaining transactional consistency. More broadly, it is important that databases combine top main memory performance (when there is enough memory) with ACID properties and the ability to operate from secondary storage as well.

\label{sec:conclusion}